\documentclass[useAMS,usenatbib]{mn2e}

\usepackage{Times}

\usepackage{amssymb,amsmath}
\usepackage{graphicx, here}
\newcommand{\gtlt}{\ensuremath{\gtrless}}         
\newcommand{\bm}[1]{\mbox{\boldmath{$#1$}}}       
\newcommand{\hs}{\ensuremath{\hspace{1em}}}       

\newcommand{\uv}[1]{\ensuremath{\hat{\bm{#1}}}}   
\newcommand{\fte}{\parbox[b]{0.4em}{\ftf \symbol{69}} \normalfont}  
\newfont{\ftf}{cmbsy12 scaled 1000}               
\newcommand{\pad}{{\small PAD}}
\newcommand{\pads}{{\small PAD }}
\newcommand{\padco}{{\small PAD}$_{co}$}
\newcommand{\padcos}{{\small PAD}$_{co}$ }
\renewcommand{\mag}{magnetosphere }
\renewcommand{\deg}{\ensuremath{^{\circ}}}        

\graphicspath{{/Users/padraig/tex/figures/}}

\title[Pulsars]{Inverse Mapping of Polarised Optical Emission from Pulsars : Basic Formulation and Determination of  Emission Altitude}
\author[J. Mc Donald]{J. Mc Donald, P. O' Connor, D. de Burca, A. Golden, A. Shearer \thanks{E-mail: andy.shearer@nuigalway.ie}\\
Centre for Astronomy, NUI, Galway, Ireland}

\begin{document}

\date{Accepted yyyy Month dd. Received yyyy Month dd; in original form yyyy Month dd}

\pagerange{\pageref{firstpage}--\pageref{lastpage}} \pubyear{2010}

\maketitle

\label{firstpage}

\begin{abstract}
We present an inverse mapping approach to determining the emission height of the optical photons from pulsars, which is directly constrained by empirical data. The model discussed is for the case of the Crab pulsar. Our method, using the optical Stokes parameters, determines the most likely geometry for emission including magnetic field inclination angle ($\alpha$), observers line of sight angle ($\chi$) and emission height. We discuss the computational implementation of the approach, along with any physical assumptions made. We find that the most likely emission altitude is at 20\% of the light cylinder radius above the stellar surface, in the open field region. We also present a general treatment of the expected polarisation from synchrotron source with a truncated power law spectrum of particles.
\end{abstract}


\section{Introduction}


Despite over forty years of theoretical and observational studies, we still do not fully understand the emission mechanism responsible for the observed pulsar radiation.

In an attempt to fully understand the high energy nonthermal emission from pulsars, both direct and indirect methods have been used. Direct models attempt to describe fully the source physics that subsequently generates the radiation we see. Conversely, inverse models use the observed radiation in an attempt to infer information about the source physics and emission geometry itself. In this paper, we present the methodology for a generic inverse method, whose aim is to restrict pulsar geometric parameters ($\alpha, \chi$, and emission location), using few assumptions and many observational constraints. Inverse models have had a long history within pulsar astrophysics. The models effectively utilise the link between lightcurve morphology and local source geometry, to place restrictions on various source parameters, i.e., to restrict the pulsar's magnetic inclination ($\alpha$), the observer's viewing angle ($\chi$), and/or to determine the emission location within the pulsar magnetosphere. Perhaps the best known of these models is the Rotating Vector Model (RVM) (\cite{RC1969}), which constrains pulsar inclination and viewing angle. It does so by assuming that the {\it radio} linear polarisation is described by a polarisation vector fixed to local B field lines, (located close to the magnetic pole), which sweeps by an observer's line of sight as the pulsar rotates.

Inverse methods in general, have evolved over time. Early ideas were to associate the width of pulse profiles with the relativistically beamed opening angle of isotropic radiation, from point sources corotating within the pulsars magnetosphere (e.g., \cite{S1970}, \cite{Z1971}, \cite{ZS1972}). \cite{M1983} proposed a model for pulsed optical $\rightarrow$ $\gamma$-ray emission based on relativistic electron beams producing highly beamed emission in the far (measured radially)  magnetosphere. Assuming emission from only the (purely dipolar) last open field lines of the {\it equator} of an orthogonal rotator, the arrival phase of emitted photons (assumed tangential to the local B, subject to aberration) were mapped onto observer phase. Aligning this radius-to-phase mapping with the relative peak arrival phases for the Vela pulsar, Morini estimated that the radio, optical and $\gamma$-ray emission for the Vela pulsar, originated at distances of $\sim 0, 0.5$ and $0.7$ $R_{LC}$ from the neutron star surface. The advent of readily available computer power allowed the construction of radius-to-phase maps, as the assumption of tangential beaming means that the mapping is independent of the specific emission process. In the late 1980s, Smith and co-workers (\cite{S1986}, \cite{SJDP1988}) extended this idea somewhat, producing radius-to-phase maps for the whole equatorial {\it plane} of an orthogonal {\it retarded} dipole. RY95 (\cite{RY1995}) succeeded in reproducing peak separations of the gamma ray pulsars, together with the radio-peak to gamma-ray peak relative phase offset, while simultaneously restricting pulsar inclination  and viewing angle. \cite{DRH2004} describe how light flight time delays can be responsible for asymmetries in pulse profiles. DR03 \citep{DR2003} developed a `Two Pole Caustic' model (TPC), which can produce double peaked (and to a more limited extent, single peak) light curves. 
While certain restrictive assumptions are sometimes necessary and justifiable, we have designed a general method tailored specifically at restricting unconstrained pulsar parameters. We effectively carry out a large scale computational search through pulsar parameter space to constrain pulsar parameters. The aim was to create a methodology of testing and refining hypotheses of how pulsars work. In this paper, we apply the model to pulsed optical emission from isolated neutron stars.

Following a successful two years of operation, the Fermi Gamma-ray observatory has identified over fifty gamma-ray pulsars (\cite{A2010b} \& \cite{SP2010}), of which 21 are radio-quiet gamma-ray pulsars and surprisingly, 11 are millisecond pulsars (\cite{A2009}, \cite{R2010}). From these observations, the polar-cap model has been effectively ruled out (\cite{A2009}, \cite{VHG2009}), although the polar cap will play a role in the initial particle acceleration. This leaves the slot-gap, outer-gap (extended outer gap for millisecond pulsars) and variations of the striped pulsar wind (\cite{AS1979}, \cite{CHR1986a}, \cite{P2009} \& \cite{KSG2002}), as the main contenders for a theory of high-energy pulsar emission.

It has long been recognised that the polarisation of pulsar radiation gives an indication of the geometry of the emission zone. \cite{TG2010} showed recently that the core and conal components of the radio emission from PSRs B1839+09, B1916+14 and B2111+46 were at low altitudes $<$ 5\% of the light cylinder radius. \cite{WW2009} mapped the magnetopshere of PSR B1055-52, indicating that the emission height was around $\sim$700 km above the neutron star's surface. In a more general way, \cite{TGG2010} have looked the geometry for radio emission assuming curvature radiation. Our objective in this paper, is to investigate the use of optical polarisation data to map the regions within the magnetosphere at which the pulsar emission can originate. The advantage of optical radiation is that its emission mechanism, incoherent synchrotron, has a simple relationship between its polarisation profile and the underlying geometry. It is also the only region where we have decent high-energy polarisation data, and such data is highly sensitive to the local emission physics conditions.

\section[]{Description of the Model}

Historically, inverse models have focused on restricting various geometrical parameters, such as pulsar inclination angle, viewing angle and/or the specific emission site responsible for the observed emission. These variables are assumed unknown initially and define a parameter search space, where the aim is to restrict them by comparison with observation. However, all such models use rather restrictive initial conditions to constrain these parameters, for example, tangential beaming from specific subsections of the magnetosphere using predefined emissivity functions. 
By removing these restrictive initial conditions and using modern compute power, we believe that the domain of applicability of inverse models can be increased. For example, we can apply the inverse approach to constrain and test emission properties, as well as geometric properties of pulsar high energy emission. Effectively, one considers the various emission parameters as being undefined and adds these parameters to the overall search space to be constrained. Extending inverse modelling in this way may provide a new way of constraining pulsar variables, and of approaching the pulsar problem in general. We describe in this section the overall approach we have designed and the specific inverse method we have developed to investigate pulsed nonthermal optical emission from pulsars.

\begin{figure}
\centering
{\includegraphics[scale=0.4]{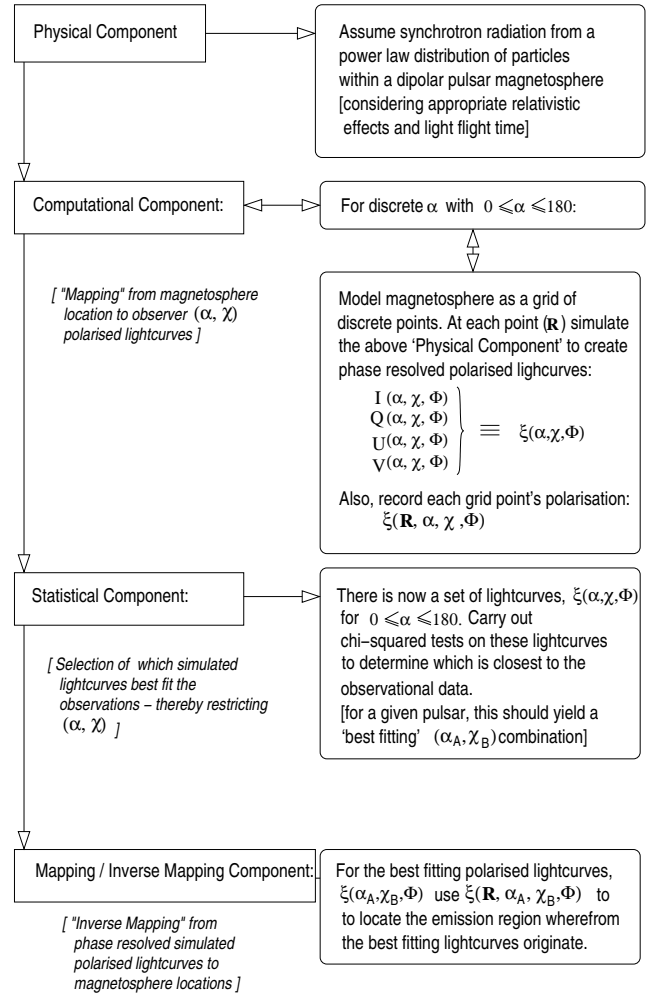}}
\caption[Representation of `Inverse Mapping'/`Search Algorithm' approach in flow diagram form]{
Representation of `Inverse Mapping'/`Search Algorithm' approach in flow diagram form. The approach
is conceptually divided into physical, computational, statistical and mapping/inverse mapping components.}
\label{flow.diagram} 
\end{figure}

Our approach is composed of a number of conceptually distinct steps, outlined in the flowchart in Figure~\ref{flow.diagram}. The main steps are subdivided into physical (P), computational (C), statistical (S), and mapping (M) elements. Briefly, these steps constitute: (1) the application of a plausible physical model responsible for the emission, (2) the computational implementation of this model, resulting in the creation of phase resolved lightcurves ($\xi(\Phi,\alpha,\chi)$)\footnote{Emission from a given pulsar is simulated and phase resolved ($\Phi$) lightcurves produced for a range of viewing angles ($\chi$). This process is repeated using different inclinations ($\alpha$) for the magnetic axis (which is not in general constrained from observations). $\xi$ can represent numerous observational properties, in this case the set of Stokes parameters I,Q,U,V.}, (3) the statistical comparison of the simulated lightcurves to observations, which results in a restriction of $(\alpha,\chi)$ parameter space enabling, (4) an `inverse mapping' into the magnetosphere, locating the regions of emission responsible the best fitting lightcurves in the first place. \\

Phase resolved Stokes parameters are produced for all possible values of $0\deg \le \alpha \le 180\deg$ and $0\deg \le \chi \le 180\deg$, in finite increments $\Delta \alpha$ and $\Delta \chi$, based on the assumption that a simple radiative emission process is occurring globally within the magnetosphere. The computational model divides the magnetosphere into a grid of points and the phase resolved angular distribution of photons from each point ($\bm{R}$) is recorded as $\xi(\bm{R},\Phi,\alpha,\chi)$. The summation of $\xi$ from all grid points creates a mapping from magnetospheric location to phase resolved observer profiles (as a function of $(\alpha,\chi)$). The statistical component selects the $\xi(\alpha,\chi)$ which best fits the observations and $\xi(\bm{R},\Phi,\alpha,\chi)$ allows an `inverse' mapping of the photon paths back into the magnetosphere.\\

These steps constitute what we refer to as the `Inverse mapping' or search algorithm approach to isolating the regions within the magnetosphere which may be responsible for pulsed optical emission. The main steps in the process are as follows: \\

\newcounter{listno}
\begin{list}{\bfseries\upshape P--\Roman{listno}}
 {\usecounter{listno}
  \setlength{\rightmargin}{0.01\columnwidth}
  \setlength{\leftmargin}{0.07\columnwidth}}

\item 
Particle motion (outside of accelerated regions) is dominated by the magnetic field, making the field an important element in any model $\Rightarrow$ we propose a magnetic field structure in the form of a retarded dipole, as the standard Deutsch formalism, \cite{ML1999}. 

\item It is plausible that synchrotron radiation is responsible for pulsed optical emission, with the observed power law spectral form indicating that the underlying particle spectrum is also of this form. Using the premise of first order approximations, together with no a-priori assumptions regarding favourable emission locations, we assume that each point in the magnetosphere can emit synchrotron radiation from an underlying particle power law, where the total number density of particles at any one point is equivalent to the Goldreich-Julian density ($n_{GJ}\propto \Omega \cdot B_z(r)$).

\item Each point $\bm{R}$ is allowed to radiate synchrotron radiation, which is characterised by the properties of the particle population at that point - essentially the energy distribution of the particles (a power law from {\bf P-II} above), and the pitch angle distribution (hereafter PAD) of the particles. The pitch angle distribution is perhaps the least well known parameter of pulsar emission models. Since the PAD is not well constrained, we assume various forms for the PAD and analyse the resultant lightcurves to choose which PADs may best represent the emission. For all PADs, we assume that the emitting particles are symmetrically distributed about the magnetic axis, with a cutoff occurring at a specific pitch angle, beyond which no emission occurs, \padco.
\end{list}

\newcounter{listno1}
\begin{list}{\bfseries\upshape C--\Roman{listno1}}
{\usecounter{listno1}
  \setlength{\rightmargin}{0.01\columnwidth}
  \setlength{\leftmargin}{0.07\columnwidth}}

\item Computationally, we represent the 3D volume of the magnetosphere as a grid. Each point on the grid is specified by local parameters such as $(\bm{R},\bm{B},n_{GJ},\bm{v}_{co})$, where $\bm{R}$, $\bm{B}$, $n_{GJ}$, and $\bm{v}_{co}$ are the spatial coordinate, the magnetic field, the local particle density and the corotational velocity respectively, at that point. The grid itself, is based on a spherical geometry, consisting of a series of fixed concentric spheres. The origin of the spheres is located at the centre of the neutron star, with points located such that $\bm{R}=(R,\theta,\phi)$ in usual spherical coordinates.

\item At each grid point ($\bm{R}$), a power law distribution of particles, having a specific PAD, emits synchrotron radiation into a certain section of sky of solid angle $d\Omega$. Relativistic aberration, beaming and doppler shifts will alter the direction of emission, the irradiated solid angle and frequency
of radiation, as seen by a sky based observer compared to an observer in the particle rest frame. These effects are considered and the emission from each point $\bm{R}$, as seen by a sky based observer, is recorded in increments of $\Delta \chi = 1\deg$, for the range of $0\deg \le \chi \le 180\deg$.

\item Since one aim of this approach is to restrict both the inclination angle $\alpha$ {\it and} the viewing angle $\chi$, step {\bf C-II} is repeated for all inclination angles with $0\deg \le \alpha \le 180\deg$, in discrete increments of $\Delta \alpha$.
\end{list}

\newcounter{listno2}
\begin{list}{\bfseries\upshape S--\Roman{listno2}}
 {\usecounter{listno2}
  \setlength{\rightmargin}{0.01\columnwidth}
  \setlength{\leftmargin}{0.07\columnwidth}}

\item Steps {\bf C-II} and {\bf C-III} create a phase space $(\alpha,\chi)$, where each point in this phase space contains phase resolved lightcurves of the form:
\begin{equation}
\xi(\bm{R}, \Phi) = \xi(R,\theta,\phi,\Phi),
\end{equation}
with $0\deg \le \chi \le 180\deg$, $0\deg \le \alpha \le 180\deg$, and where  $\xi$ represents each of the four Stokes parameters $(I,Q,U,V)$. A $\chi$-squared goodness of fit test is performed between each of the simulated Stokes parameters and the observed Stokes parameters, at each point in $(\alpha,\chi)$ phase space, to produce the best fitting $(\alpha,\chi)$ combination.
\end{list}

\newcounter{listno3}
\begin{list}{\bfseries\upshape M--\Roman{listno3}}
 {\usecounter{listno3}
  \setlength{\rightmargin}{0.01\columnwidth}
  \setlength{\leftmargin}{0.07\columnwidth}}

\item Knowing the best fitting $(\alpha,\chi)$ combination, $\xi(R,\theta,\phi,\Phi,\alpha,\chi)$ is used to inverse map the constituent photons back into the magnetosphere, thereby locating the origin of the emission, which creates the best fitting lightcurves ($\xi$). The visualisation of these regions will hopefully yield an indication of the originating locations.
\end{list}

In the following sections, each of these steps in illustrated in more detail, applying the search algorithm to a simulated pulsar with Crab like parameters: ($P=33$ ms, $\dot{P}=4.209\times10^{-13}\rm{ss^{-1}}$ and $B_{\rm{surf}}=3.8 \times 10^{12}$ G).

\section[]{Description of the Physical Model}

It is assumed in this modelling that at any point in the magnetosphere, radiation arises through synchrotron radiation from a power law spectrum of particles: the particle index, p, is fixed via the observed photon spectral index s, through the relation, s=(p+1)/2, where:

\begin{equation}
F_\nu \propto \nu^{-s} \quad \Rightarrow \quad N(E) \propto E^{-p}.
\end{equation}

When simulating emission over a range of frequencies (e.\,g., for U, B, V bands), contributions are summed only from those particles whose energies, are such that they contribute significantly to emission at the {\it observed} frequency $\nu$.

\subsection{Synchrotron Radiation}

Synchrotron radiation from a single particle, spiralling around a local magnetic field ($\bmath B$), at a pitch angle $\alpha$ (where $\hat{n} \cdot \hat{B} = \cos \alpha$ with $\hat{n}$ the instantaneous velocity of the particle), produces a continuum spectrum of frequencies with a peak emissivity close to a critical frequency, $f_c$ where:

\begin{equation}
f_c(E,B,\theta) = \frac{3}{2} f_B \sin \theta \left( \frac{E}{E_o} \right)^2,
\label{fc}
\end{equation}

with $f_B=e B/2\pi m$, being the particle cyclotron frequency. To first order, the spectrum exhibits a quadratic rise and exponential drop as the frequency passes through the critical value. At lower frequencies, the spectrum consists of discrete harmonics of the fundamental (equation~\ref{fundamental_freq}) frequency below which no emission is possible.

\begin{equation}
f_f = \frac{f_B}{\sin^2 \theta} \frac{E_o}{E}  \label{fundamental_freq}.
\end{equation}

\begin{figure}
\center
{\includegraphics[scale=0.4]{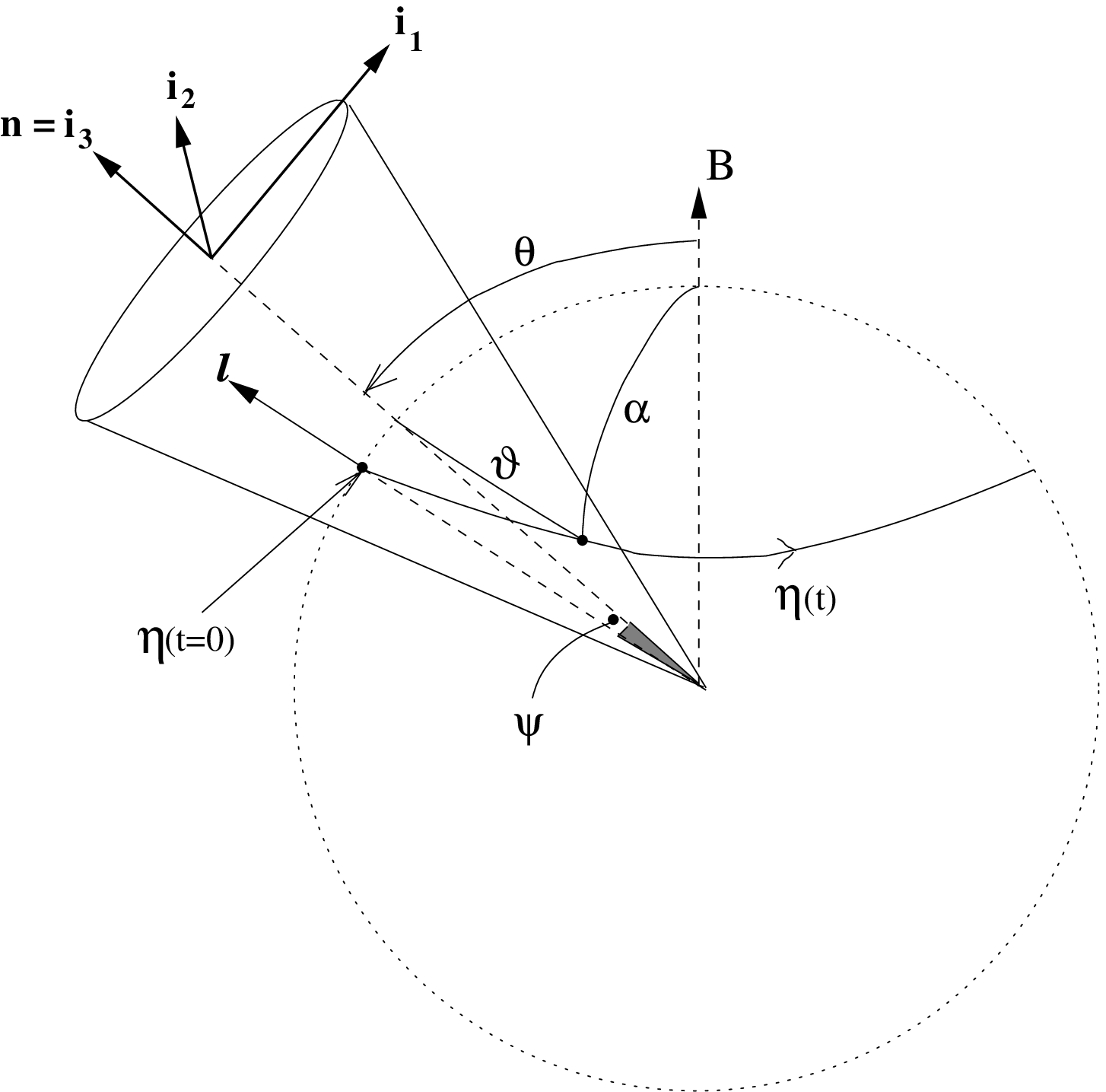}} 
\caption{Important source and observer geometric parameters relating to synchrotron emission.}
\label{particle_geometry}
\end{figure}

The emitted radiation is in general, elliptically polarised with the principle axes of the polarisation ellipse aligned parallel and perpendicular to the projection of $\bmath{B}$ on the plane transverse to the emission direction, $\hat{l}$ (see Figure~\ref{particle_geometry}). The angular dependence of the polarised emissivity depends strongly on the angle between the line of sight ($\hat{l}$) and $\hat{n}$, where we write, $\hat{n} \cdot \hat{l} = \cos \psi$ with $sgn(\psi)=\alpha-\theta$. The case $\psi = 0$, results in linear polarisation perpendicular to $\bmath{B}$, $\psi > 0$ denotes elliptical polarisation having the major axis perpendicular to the projection of $\bmath{B}$, and $\psi < 0$ indicates right handed polarisation and $\psi > 0$ left handed.

For a population of particles, one must integrate emission over the velocity distribution, which amounts to an integration over the pitch angle distribution and over the range of particle energies present. Results for a power law spectrum of particles ($N(E) \propto E^{-\gamma}$) are well known; the emitted radiation spectrum is also a power law, with the photon spectral index related to the particle index, where $\alpha=(\gamma+1)/2$ and $\gamma < 1/3$, in order to maintain a finite integral. This monotonic spectrum is of course a special case, which results from integrating over a population having infinitely wide energy bounds. So that practically, the formalism can be applied only at frequencies which are unaffected by any issues related to the nature of finite particle energy bounds. 

\subsection{A Truncated Power Law Spectrum of Particles}

In any real situation, one is dealing with particles having finite upper and lower energy bounds, which alter the spectrum significantly from the classical monotonic scenario. For the purposes of this work, the angle dependant and integrated polarisation properties of synchrotron emission from a truncated power law spectrum of particles, with $N(E) \propto E^{-\gamma}$ is considered. In this spectrum, any given particle's energy is confined between lower ($E_1$) and upper ($E_2$) limits, such that $E_1 \le E \le E_2$ having a (potentially) isotropic, axially symmetric (about $\bm{B}$) pitch angle distribution. Such a distribution is rather general and has been dealt with in detail by GLW74 \citep{GLW1974}. They derive emissivity properties using the polarisation tensor for light, which represents the cross correlated quadratic components of the electromagnetic wave field as a rank 2 tensor, $\rho_{\alpha\beta}$ (where $\alpha$ and $\beta$ represent the component directions in the transverse plane). The tensor $\rho_{\alpha\beta}$, contains the complete polarisation content of the radiation and can therefore be related to the more commonly used Stokes parameters, as illustrated in equation~\ref{polarisation_tensor}. For reference purposes, the underlying polarised emissivity properties are given here as equations~\ref{stokes}, reproduced from GLW74.

\begin{equation}
\left.
\parbox{0.5\textwidth}{
\begin{eqnarray*}
I & = & \frac{A\mu e^2c}{2\sqrt{2}}(\frac{3}{2})^{\gamma/2}\phi(\theta)({\nu}_H \rm{sin} \theta)^{(\gamma+1)/2} \\
 	                                   &  &\cdot\:\: \nu^{-(\gamma-1)/2} \left[ \mathcal{J}_{(\gamma+1)/2} \right]_{x_2}^{x_1} \nonumber  \\ 
Q & = & \frac{A\mu e^2c}{2\sqrt{2}}(\frac{3}{2})^{\gamma/2}\phi(\theta)
            ({\nu}_H \rm{sin}\theta)^{(\gamma+1)/2} \\
            &  & \cdot\:\: \nu^{-(\gamma-1)/2}\left[ \mathcal{L}_{(\gamma+1)/2} \right]_{x_2}^{x_1}  \nonumber \\
U & = & 0 \nonumber  \\
V & = & \frac{A\mu e^2c}{\sqrt{3}}(\frac{3}{2})^{\gamma/2}\phi(\theta)
                            \rm{cot} \theta ({\nu}_H \rm{sin}\theta)^{\gamma/2+1}
                             \nu^{-(\gamma/2)} \nonumber  \\
                           &  & \times \left[\mathcal{R}_{(\gamma/2+1)}+
                                 (1+g(\theta))(\mathcal{L}_{\gamma/2} -\frac{1}{2}\mathcal{J}_{\gamma/2}) 
                                 \right]_{x_2}^{x_1} \label{stokes}
\end{eqnarray*}} 
\right.        
\end{equation}
where:
\begin{equation}
\mathcal{J}_{n}(x)=\int_o^x \xi^{n-2} F(\xi) d\xi, \hs \hs \mathcal{L}_{n}(x)=\int_o^x \xi^{n-2} F_p(\xi) d\xi
\end{equation}
\begin{equation}
\mathcal{R}_{n}(x)=\int_o^x \xi^{n-2} F_s(\xi) d\xi, \hs \hs F(\xi)= \int_\xi^\infty K_{5/3}(y)dy
\end{equation}
\begin{equation}
F_p(\xi)= \xi K_{2/3}(\xi), \hs \hs F_s(\xi)= \xi K_{1/3}(\xi)
\end{equation}
so that:
\begin{equation}
\mathcal{J}_{n} \equiv \int_o^\infty \xi^{n-2} F(\xi) d\xi  \hs \rm{etc.,}
\end{equation}
where the $K_\alpha$, are modified Bessel functions and the variable $x \equiv f/f_c$, where $f_c$ is the critical frequency given by \ref{fc}:

The effect of the $truncated$ power law energy spectrum is seen in the fact that the functions modifying the observed Stokes parameters (i.e., $\mathcal{J}_n(x), \mathcal{L}_n(x)$ and $\mathcal{R}_n(x)$), must be evaluated between upper ($x_1$) and lower ($x_2$) limits of the parameter `x', where $x=x(B,\theta,E)$. Each polarised emissivity parameter is dependant on a functional combination of $\phi(\theta), f_B \sin \theta, f, \gamma$, and this functionality is modulated by finite integrals over the particle population (having integrand `x').

\begin{figure}
\center
{\includegraphics[scale=0.4]{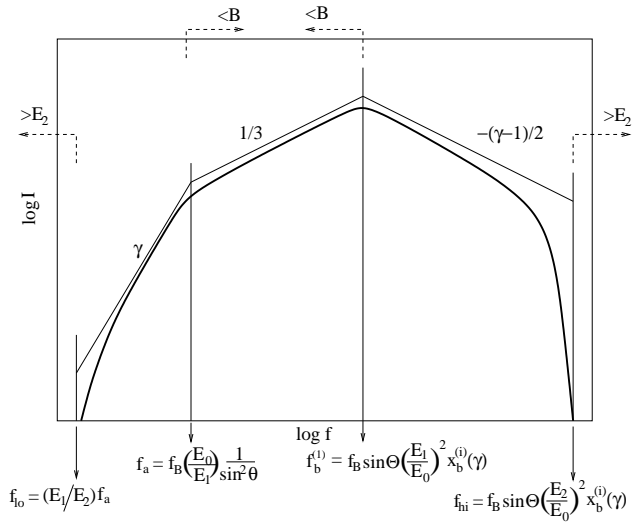}} 
\caption{Spectral variation of emission (Stokes I) from a truncated power law energy spectrum. The effect that decreasing the local magnetic field strength ($\bm{B}$) and increasing the upper power law cutoff energy ($E_2$) have on the spectral break points is indicated by the directional dotted arrows at the appropriate points. Our spectral range covers UBV.
}
\label{spectrum_with_corrxn_factors}
\end{figure}

As analysed in detail by GLW, having a finite particle energy range results in an emission spectrum with lower ($f_l$) and upper ($f_u$) frequency bounds. This spectrum possesses roughly two spectral breaks at frequencies $f_a$ and $f^{(i)}_b$, where $f_{lo}<f_a<f^{(i)}_b<f_{hi}$, with differing spectral indices for each polarisation parameter, each asymptotically dependent on the power law index $\gamma$ (see for example, Figure~\ref{spectrum_with_corrxn_factors}).          \\

The frequency bounds, the internal transition frequencies and the form of the spectral variation, are all the result of energy cut-off effects. All effects have a simple physical basis and can be described in terms of  the very useful variable of integration `x', which contains the frequency of interest, f, and encodes all relevant physical factors relating to the emission process, ($E, {\bf B}, \theta$). In fact, the fiducial frequencies are derived from the corresponding fiducial `x-factors', designated $x_l<x_a<x^{(i)}_b<x_u$.

Physically, both the low and high energy frequency cutoffs are due to the absence of particles with energies, $E$, greater than the upper limit $E_2$ ($f_{lo} \propto E_2^{-1}$, $f_c \propto E_2^2$). Above $x^{(i)}_b$, we have $x_1 \ll x \ll x_2$, so that the upper and lower limits of integration can be approximated to $\infty$ and 0 respectively. This corresponds to the classical scenario in which $f_{c1} \ll f \ll f_{c2}$; the case in which the integrals are dependant solely on the spectral index $\gamma$, and can be expressed in terms of a combination of Euler Gamma functions. The photon indices for the stokes parameters are now the classical values, the frequency dependencies already given in relations \ref{stokes}, i.e., $\alpha=(\gamma-1)/2$ for Stokes I and Q, and $\alpha=\gamma/2$ for Stokes V. Below $x^{(i)}_b$, the finite limits of $x_1$ and $x_2$ cannot be ignored, and one gets a `falling away' from the classical photon index as the frequency decreases. This results in an explicit dependence on $x_1$ and $x_2$ (and subsequently an altered frequency dependence). Deriving Power Law Expansion (PLE) approximations for the relevant integral evaluations, GLW74 showed that the asymptotic spectrum should now vary as $(f/f_B \sin \theta)^{1/3} E^{\gamma-1/3}$ for Stokes I and Q, and as $E^{-\gamma}$ for Stokes V (where $E=E_1$ or $E_2$).

At $f^{(i)}_b$, again, the finite limits of $x_1$ and $x_2$ cannot be ignored and one gets a `falling away' from the classical photon index as frequency decreases. Power law approximations to the integrals in relations \ref{stokes}, show an asymptotic dependence on $x_1$, providing a spectral variation of $f^{1/3}$.

As lower energy particles will have a higher fundamental frequency than higher energy particles, a point may come where some of the lower energy particles in the population cannot contribute to the frequency of interest, and the frequency at which this occurs, is designated $f_a$. GLW74 deal with this case by determining the lowest energy particles from the population which can contribute to emission at frequency f (designated by $E_1^{'}$ where $E_1^{'} > E_1$), and replacing the upper limit of integration with $x_1^{'}$, where $x_1^{'}=f/f_c(E_1^{'})$. It is this replacement, which results in the augmented spectral index of $\alpha=-\gamma$ at frequencies below $f_a$.

The different energy cut-off effects delineate different regions in which different physical effects alter the emission characteristics. The effects are encoded in the integration x-factors, both in the form of $x_l$ and the extent of the range $(x_l,x_2)$, over which emission is allowed. Figure~\ref{spectrum_with_corrxn_factors} shows the expected form of the emissivity for a truncated power law particle population emitting synchrotron radiation.\footnote{We note that the photon indices quoted above are asymptotic values evaluated in the limit of `small', using power law expansion representations of the functions $\mathcal{J}_{n}(x), \mathcal{L}_{n}(x)$ and $\mathcal{R}_{n}(x)$ derived by GLW74. The transition frequencies and photon indices are therefore asymptotic and indicative of the actual spectrum, which should transit smoothly through the transition frequencies and deviate smoothly from the asymptotic spectral indices.}

In our approach to modelling emission from a truncated power law particle population, we follow the GLW74 approach and describe the spectral variability in terms of correction factors, $C^i$, $i=1,2,4$ applied to the standard power law dependences (see Appendix \ref{appendix_1}). We note that in limited circumstances, it is possible for $x_a>x^{(i)}_b$, so that the spectral emissivity has a lower and upper bound with a single internal transition frequency. This situation would arise when the upper and lower bounds are sufficiently close together. For completeness, we include the relevant correction factors in Appendix \ref{appendix_1} (as this situation was not discussed in GLW74).

\subsection{Effect of source motion}

The description of the polarisation parameters above is valid when the magnetic field is stationary relative to an observer. In the case of emission from synchrotron radiating particles constrained to move along magnetic field lines in a pulsar magnetosphere, the underlying field structure is (relativistically) rotating. To describe the effect of the source motion on the observed polarised emission, we consider how a general Lorentz boost transforms the radiation fields. Also, to extract angular dependences and for frame transformations, it is convenient to use the polarisation tensor, $\rho_{\alpha\beta}$, representation of light.   

In our notation, a noninertial observer  in the corotating frame $S$ (basis $O_{xyz}$), sees a stationary magnetic field $\bm{B}$ and views this field along a direction $\bm{\hat{n}}$, such that $\bm{B}_f \cdot \bm{\hat{n}} \propto  \cos \psi$. The observer sees light from a particle, if $\bm{\hat{n}}$ is sufficiently close to $\bm{\tau}(t)$, where $\bm{\tau}$ describes the trajectory of the charged particle (see Figure~\ref{particle_geometry}). The electromagnetic field of the light oscillates in a plane transverse to $\bm{\hat{n}}$ (the `observer plane' K), and may be expressed in component form along two mutually perpendicular axes $(\bm{\hat{i}}_1, \bm{\hat{i}}_2)$ in the plane K. Frame $S'$ observes the magnetic field $\bm{B}_f$ to move at an arbitrary constant velocity $\bm{\beta}=\bm{v}/c$. An observer in $S'$ will therefore see light emitted at an aberrated direction $\bm{\hat{n}}'$ about the boosted field direction $\bm{B}'$. The components of the electromagnetic wave will also be boosted ($\bm{E}',\bm{B}'$). A general Lorentz boost, where frame $S^{\prime}$ moves at an arbitrary constant velocity $\bm{\beta}$ to frame $S$, is given as $\Lambda^\mu_\nu$ (equation~\ref{lorentzboost}) where a four-vector $\vec{x}$ transforms as ${x'}^\mu = {\Lambda^{\mu}}_\nu x^\nu$.

\begin{equation}
\label{lorentzboost}
\Lambda^\mu_\nu = \left[ \begin{array}{llll}
  \gamma        & -\beta_x\gamma  & -\beta_y\gamma & -\beta_z\gamma \\
 -\gamma\beta_x & 1+\frac{\gamma-1}{\beta^2}{\beta_x}^2 & \frac{\gamma-1}{\beta^2}\beta_x\beta_y & \frac{\gamma-1}{\beta^2}\beta_z\beta_x \\
 -\gamma\beta_y & \frac{\gamma-1}{\beta^2}\beta_x\beta_y & 1+\frac{\gamma-1}{\beta^2}{\beta_y}^2 & \frac{\gamma-1}{\beta^2}\beta_z\beta_y \\
 -\gamma\beta_z & \frac{\gamma-1}{\beta^2}\beta_x\beta_z & \frac{\gamma-1}{\beta^2}\beta_y\beta_z & 1+\frac{\gamma-1}{\beta^2}{\beta_z}^2\\
\end{array} \right] 
\end{equation}

\noindent The electric and magnetic fields, transform as the components of the antisymmetric tensor $F_{\mu\nu}$, (equation~\ref{antisym_eb}), so that ${F'}_{ik}={\Lambda^{\beta}}_k {\Lambda^{\alpha}}_iF^{\alpha \beta}$. The transformed wave four-vector ($\bm{k'}$), electric ($\bm{E'}$) and magnetic ($\bm{B'}$) fields are given in equations~\ref{E_trans} $-$ \ref{k_trans}. 

\begin{equation}
F_{\mu\nu} = \left[ \begin{array}{cccc} 
      0   & -E_x & -E_y & -E_z \\
      E_x & 0    & B_z  & -B_y \\
      E_y & -B_z & 0    & B_x \\
      E_z &  B_y & B_x  & 0   
      \label{antisym_eb}
      \end{array} \right]
\end{equation} 
     
\begin{eqnarray}
\bm{E}' & = & \gamma(\bm{E} + \left[ \bm{\beta H} \right]) 
       - \frac {\gamma^2}{(\gamma+1)}\bm{\beta} (\bm{\beta E}) \label{E_trans}\\
\bm{B}' & = & \gamma(\bm{B} + \left[ \bm{\beta B} \right]) 
       - \frac{\gamma^2}{(\gamma+1)}\bm{\beta} (\bm{\beta B}) \label{B_trans}
\end{eqnarray}

\begin{equation}
{k'}_i = \frac{\omega}{c}\gamma(1-\bm{\beta}\bm{n})
         \left\{ 
         1,\left( \frac{\bm{\beta-\hat{n}}}{1-(\bm{\beta}\bm{n})}-
                  \frac{1}{(\gamma+1)}\frac{\bm{\beta}(\bm{\beta n})}{1-(\bm{\beta}\bm{n})}
           \right) \right\} \nonumber
\end{equation}

\begin{equation}
= \left( \frac{{\omega}'}{c},\frac{-{\omega}'}{c}\bm{\hat{n}'} \right) \label{k_trans}
\end{equation}

It is now necessary to describe the observed intensity and polarisation in terms of quantities seen in the observers frame (S'). For this, the polarisation tensor representation ($\rho_{\alpha\beta}$) of light is most useful, given by:
\begin{equation}
\label{polarisation_tensor1}
\rho_{\alpha \beta} = \frac{cR^2}{4\pi}\fte_{\,\,\alpha} {\fte}^{\,\,\,*}_{\,\,\beta},
\end{equation}
\noindent where $\fte_{\,\,\alpha}$ and $\fte_{\,\,\beta}$ are two orthogonal components of the electromagnetic wave field. The polarisation tensor can also be expressed in terms of the more widely used Stokes parameters, as in equation~\ref{polarisation_tensor}.

\begin{equation}
\label{polarisation_tensor}
\rho = \left[ \begin{array}{cc} 
      \rho_{11} & \rho_{12} \\
      \rho_{21} & \rho_{22} \\
      \end{array} \right] \propto 
      \left[ \begin{array}{cc} 
      \frac{1}{2}(I+Q) & \frac{1}{2}(U-iV) \\
      \frac{1}{2}(U+iC) & \frac{1}{2}(I-Q) \\
      \end{array} \right]  
\end{equation}

\begin{figure*}
\center
{\includegraphics[scale=0.6]{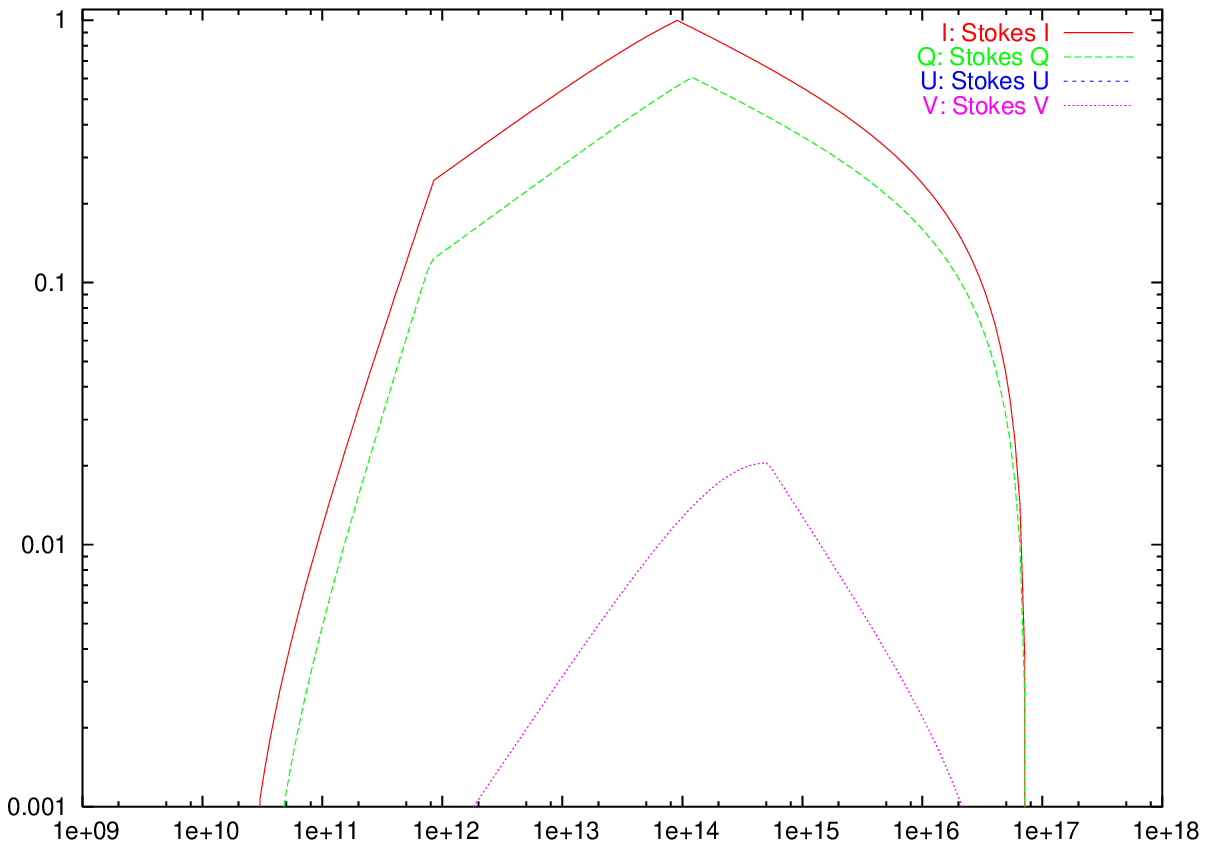}
\hfill
\includegraphics[scale=0.6]{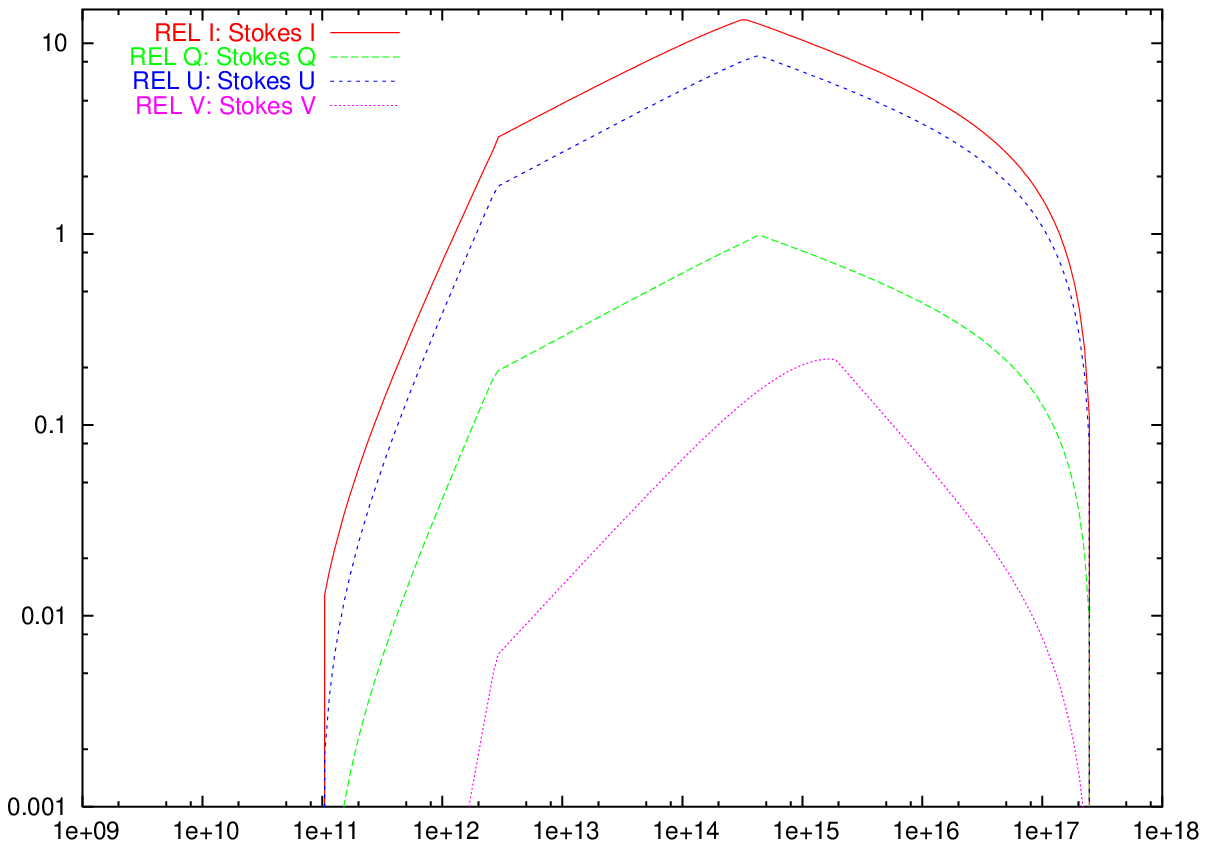}} 
\caption{Effect of a relativistic source motion on the observed Stokes parameters, determined here for $B=1.9 \times 10^{6}$ G, $\beta=0.87$.}
\label{boosting_on_stokes}
\end{figure*}

Since the fields of the wave remain transverse in any frame, it is apparent that the tensor $\rho_{\alpha\beta}$ will remain two dimensional in any new frame. Also, because the transformation is real, the real and imaginary parts of the tensor transform independently. These properties allow us to boost and rotate the electric field components from plane K in frame S, to plane K' in frame S'. Before boosting $\rho_{\alpha\beta}$, any reference to $\bm{B}$ in the expression for $\bm{E'}$ is removed, using the substitution $\bm{B}=[\uv{n}\rm{E}]$. Also, a right handed orthonormal basis set (denoted $\rm{O}_{j_1j_2j_3}$) is defined in the observers (S') plane (K'). The projection of the pulsars rotation axis ($\bm{\Omega}$) on the plane K' is chosen as a uniform reference direction $\uv{j_1}$, so that the basis set designated as $\rm{O}_{j_1j_2j_3}$, can be defined through equation~\ref{j_basis}.

\begin{equation}
\bm{j}_1=\frac{[\bm{n}'\uv{\Omega}]}{\mid[\bm{n}'\uv{\Omega}]\mid} ; \quad \bm{j}_2=[\bm{n}'\bm{j}_1]
               ; \quad \bm{j}_3=\bm{n}' \label{j_basis}
\end{equation}

$\uv{j_1}$ and $\uv{j_2}$ are orthogonal and in the observers plane (K'), perpendicular to the boosted emission direction $\uv{j_3}$. We now express  $\bm{E'}  =  \bm{E_1}'\uv{j_1} + \bm{E_2}'\uv{j_2}$, where $\bm{E_1}' = (\bm{E}' \uv{j_1})$ and  $\bm{E_2}' = (\bm{E}' \uv{j_2})$. Since $\bm{E}= iE_1 \uv{i_1} + E_2 \uv{i_2}$, it is possible to write $\bm{E}'$ in the $\rm{O}_{j_1j_2}$ basis explicitly in terms of $\bm{E_1}$ and $\bm{E_2}$ as follows:

\begin{eqnarray}
({E}_1'\bm{j}_1, \, {E}_2'\bm{j}_2) & = & (\{iE_1x_{11}+E_2x_{21}\}\bm{j}_1, \, \nonumber \\
  &  & \{iE_1x_{12}+E_2x_{22}\}\bm{j}_2),
\label{transformed}
\end{eqnarray}

where the x factors ($x_{11}$ etc.), are given by equation~\ref{transformation_factors}, and are dot products between the $\rm{O}_{j_1j_2}$ basis vectors and the $E_1$ and $E_2$ components of $\bm{E}'$.

\begin{eqnarray}
  x_{\alpha\beta}=\frac{1}{(1+(\bm{\beta}\bm{n}'))}\Bigg\{ \frac{\bm{l}_\alpha\bm{j}_\beta}{\gamma}+ \nonumber \:\:\:\:\:\:\:\:\:\:\:\:\:\:\:\:\:\:\:\:\:\:\:\:\:\:\:\:\:\:\:\:\:\:\:\: \\
  \:\:\:\:\:\:\:\:\:\:\:\:\:\:\:\:(\bm{\beta}\bm{l}_\alpha)\left[ (\bm{n}'\bm{j}_\beta) +\gamma(\bm{\beta}\bm{j}_\beta)\left( 1+\frac{\gamma}{(\gamma+1)}(\bm{\beta}\bm{n}') \right) \right] \Bigg\} \label{transformation_factors}
\end{eqnarray}

Equation~\ref{transformed} expresses how the components of $\bm{E}$ and $\bm{B}$ in plane K, written in terms of the basis ($\bm{\hat{i}}_1, \bm{\hat{i}}2$), transform into components of $\bm{E}'$ and $\bm{B}'$ in plane $K'$, expressed along a basis $\bm{\hat{j}}_1, \bm{\hat{j}}_2$. Using equation~\ref{polarisation_tensor}, the components of the emission-polarisation tensor ${\rho'}_{\alpha \beta}$ in frame $S'$ can be determined as follows:

\begin{equation}
 \begin{array}{lll} 
\rho_{11} = \fte_{\,\,\,1}{\fte}^{\,\,\,\,*}_{\,\,\,1}x_{11}x^{*}_{22} + \fte_{\,\,\,2}{\fte}^{\,\,\,*}_{\,\,\,2}x_{21}x^*_{21}, \:\:\:\:\:\:\:\:\:\:\:\:\:\:\:\:\:\:\:\: \\
\\
\rho_{12} = \fte_{\,\,\,2}{\fte}^{\,\,\,\,*}_{\,\,\,2}x_{21}x^*_{22} + \fte_{\,\,\,1}{\fte}^{\,\,\,*}_{\,\,\,1}x_{11}x^*_{12} \:\:\:\:\:\:\:\:\:\:\:\:\:\:\:\:\:\:\:\: \\
\:\:\:\:\:\:\:\:\:\:\:\: +i(\fte_{\,\,\,1}{\fte}^{\,\,\,*}_{\,\,\,2}x_{11}x^*_{22} - \fte_{\,\,\,2}{\fte}^{\,\,\,*}_{\,\,\,1}x_{21}x^*_{12}), \\
\\
\rho_{21} = \fte_{\,\,\,2}{\fte}^{\,\,\,\,*}_{\,\,\,2}x^*_{21}x_{22} + \fte_{\,\,\,1}{\fte}^{\,\,\,*}_{\,\,\,1}x^*_{11}x_{12} \:\:\:\:\:\:\:\:\:\:\:\:\:\:\:\:\:\:\:\: \\
\:\:\:\:\:\:\:\:\:\:\:\: +i(\fte_{\,\,\,1}^{\,\,\,*}{\fte}_{\,\,\,2}x_{11}^*x_{22} - \fte_{\,\,\,2}^{\,\,\,*}{\fte}_{\,\,\,1}x^*_{21}x_{12}), \:\:\:\:\:\:\:\:\:\:\:\:\:\:\:\:\:\:\:\: \\
\\
\rho_{22} = \fte_{\,\,\,2}{\fte}^{\,\,\,\,*}_{\,\,\,2}x_{22}x^*_{22} + \fte_{\,\,\,1}{\fte}^{\,\,\,*}_{\,\,\,1}x_{12}x^*_{12}.
  \end{array} 
\end{equation}

The polarisation tensor can be related to the Stokes parameters in the standard way (see equation~\ref{polarisation_tensor}), and using these relations, the Stokes parameters observed in frame $S'$, can be expressed in terms of the field components of ($\bm{E}, \bm{B}$), as calculated in frame $S$. In the same way, $(I',Q',U',V')$ in $S'$ can be related to $(I,Q,U,V)$ in frame $S$. Carrying out the algebra the results are as follows:

\noindent
\parbox{0.8\columnwidth}{
\begin{eqnarray}
I' & = &   \frac{1}{2} \left\{ (x_{11}^2+x_{12}^2)(I+Q)+(x_{21}^2+x_{22}^2)(I-Q) \right\}  \nonumber \\
Q' & = &   \frac{1}{2} \left\{ (x_{11}^2-x_{12}^2)(I+Q)-(x_{22}^2-x_{21}^2)(I-Q) \right\}  \nonumber \\ 
U' & = &   \hspace{1em}\frac{}{} (x_{12}x_{11})(I+Q)+(x_{22}x_{21})(I-Q) \nonumber \\ 
V' & = &   \hspace{1.3em}(x_{22}x_{11}-x_{12}x_{21}) \, V \nonumber 
\end{eqnarray}}
\hspace{0.3cm}
\parbox{0.05\columnwidth}{$\left.
  \begin{array}{c}   
               \\ \\ \\  \\ \\ \\ \\                 
  \end{array}
  \right\} $} 
\hfill 
\parbox{0.05\columnwidth}{\begin{eqnarray}\label{test-label} \end{eqnarray}}

A significant source motion will alter the observed polarisation, the intensity and magnitude of the Stokes parameters will change, and this will be accompanied by a Doppler shift in the observed frequency. The appearance of a non-zero Stokes U in the observer frame is not in itself significant, as the relative magnitude of Stokes Q and U, depend on the orientation of your reference axes in the observer plane. Figure~\ref{boosting_on_stokes} shows the source and observer polarisation spectrum for a source magnetic field of $1.9\times10^6$ G, with a source velocity of $\beta=0.87$.

\section[4]{Description of the Computational Model}
\label{sec:4}

The above physical emission model is simulated within the pulsar's magnetospheric environment. The three dimensional magnetosphere is represented computationally in spherical polar coordinates, and it is assumed that emission arises from within the open volume of the magnetosphere, which is determined numerically. As such, emission is only simulated from those grid points lying in the open volume, where the grid itself is specified using coordinates $(r,\theta,\phi)$. The number of grid points on any given spherical surface is controlled by the parameters $N_\theta$ and $N_\phi$, (the number of $\theta$ and $\phi$ divisions). The parameter $N_r$ specifies the number of divisions in the radial direction, and the grid itself is devised in such a way that each grid point represents an equal volume of space, so that in essence, the discretisation of the magnetosphere is homogeneous with respect to volume. 

The emission is calculated at all points in the open volume, which requires the determination of {\it local} parameters. These include the magnetic field strength, $\bm{B}$, the total number of particles at a specific point, $n_{GJ}$, the assumed pitch angle distribution, local co-rotational velocity and the range of particle gamma factors which contribute to observable emission at that particular point. All necessary dependencies are calculated at each grid point to create the local emission profile, which effectively determines how the polarised spectral emissivity varies across the extent of the local pitch angle distribution. Each profile is subject to the required frame transformations (described above), which determine the emissivity profile as seen by a stationary observer external to the rotating magnetosphere.

\begin{figure*}
\center
{\includegraphics[scale=0.4]{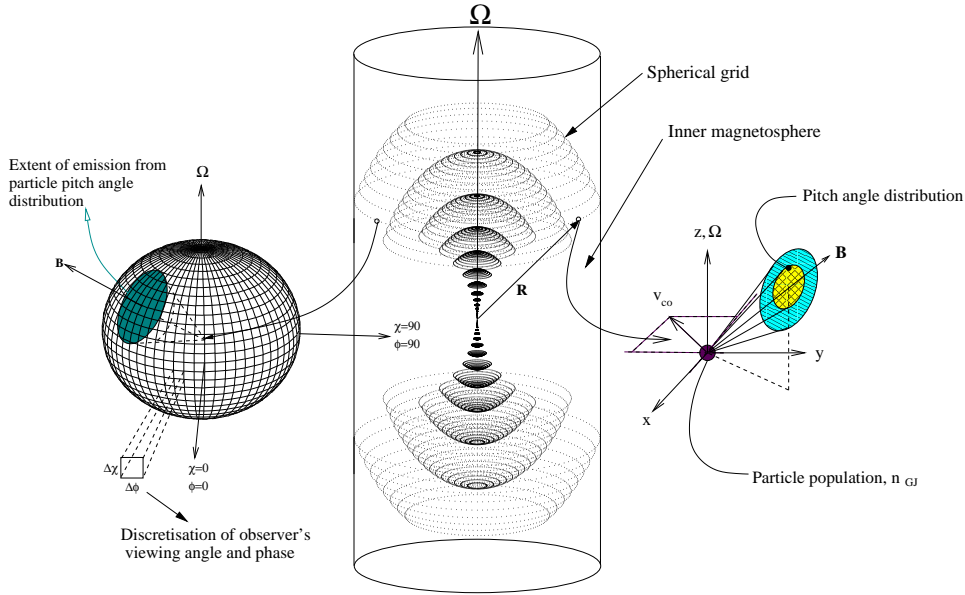}} 
\caption{Centre: The concentric spherical Cartesian grid imposed on the pulsar magnetosphere, showing points only in the open volume of an aligned rotator ($\uv{\Omega}\cdot\uv{\mu}=1$). Right: At each individual location ($\bm{R}$), we calculate the radiation emitted from the local particle distribution, taking the local physical conditions into account - e.g., $\bm{B}, \bm{v}_{co}$, the power law particle distribution, the specific pitch angle distribution etc. Left: Representation of how emission from a particle pitch angle distribution extends over a range of viewing angles ($\chi$) and phase ($\Phi$). Emission is recorded computationally into discrete bins of extent $\Delta\chi \times \Delta\Phi$ (as described in section \ref{sec:4}).}
\label{spherical_grid}
\end{figure*}

The pitch angle distribution at any open volume point, generates emission which extends over a finite solid angle. This solid angle of emission has a finite extent latitudinally and longitudinally with respect to an observer, which directly correspond to a range in observer viewing-angle, $\chi$ and temporal phase, $\Phi$, respectively. Discretising this observer space\footnote{Representing $0\deg \le \chi \le 180\deg$ and $0 \le \Phi \le 1$ with step-sizes of $\Delta \chi$ and $\Delta \Phi$.} allows emission from any point to be recorded as a function of viewing angle and phase. For any given model pulsar (i.e., for a specific magnetic inclination, $\alpha$), the code records the phase resolved polarimetry in the form of an array, denoted as $\xi(R,\theta,\phi,\Phi,\alpha,\chi)$ ($=\xi(\bm{R},\Phi,\alpha,\chi)$). This means that for each grid point location, $\bm{R}$, the intensity and phase resolved Stokes parameters for the range of viewing angles over which emission is seen, are recorded.\footnote{The magnetosphere is of course rotating, so at any point in time, the correct phase of emission is determined taking light flight time, aberration and rotation into account.}\\

The indices in the $\xi$ array are discrete and represents a record of the phase resolved Stokes parameters seen by observers from point $\bm{R}$. Once all grid points have been sampled, the array $\xi$ can be summed from all appropriate locations to form $\xi(\alpha, \chi, \Phi)$\footnote{Represented in Figure~\ref{flow.diagram} as the array $\xi(\alpha_A,\chi_B,\Phi)$.} - i.e., the phase resolved Stokes parameters observed from the entire magnetosphere for a pulsar inclination of $\alpha$. \\

A single run of this code will create files which contain $\xi(\alpha,\chi,\Phi)$ for a single, specific pulsar inclination $\alpha$. The search algorithm approach attempts to restrict $\alpha$ and $\chi$ by simulating emission for the range of all {\it possible} $\alpha$ values\footnote{By explicit assumption, there is no a priori restriction on $\alpha$ (which is generally the case, as observations constrain neither $\alpha$ nor $\chi$).} ($0\deg \le \alpha \le 180\deg$), and subsequently selecting those simulated lightcurves which best fit the observational data (which should correspond to a unique combination of $\alpha$ and $\chi$). To generate lightcurves for different possible inclinations of a given pulsar,  $\alpha$ is varied from $0\deg \le \alpha \le 180\deg$ in discrete steps $\Delta \alpha$. This generates a large pool of phase resolved model lightcurves, each dependant on a unique combination of $\alpha$ and $\chi$ - for example, with $\Delta \alpha = 10\deg$ and $\Delta\chi=1\deg$, one generates $180\times18= 3240$ distinct lightcurves ($\times 4$ when polarimetry is considered).

\section{Results}

The search algorithm approach was partly motivated on the premise that any simulated Stokes parameters $(\xi)$, would be a variable function of the phase space $(\alpha,\chi,$ \pad, \padco). The simulated Stokes parameters vary as a smooth function of all phase space variables $\xi(\alpha,\chi,$ \pad, \padco\footnote{PAD refers to the pitch angle distribution which can either be isotropic or have a Gaussian structure; \padco defines the maximum  pitch angle.}  ), thereby allowing for the selection of a best fitting parameter set.
This being the case, one can restrict the pulsar geometry ($(\alpha,\chi)$ combination), which best fits observations and subsequently carry out the inverse mapping step to locate the associated source region within the pulsar magnetosphere.\\

Simulated Stokes parameters are by definition dependent on all phase space variables $(\alpha,\chi,$ \pad,\padco). As described in section \ref{sec:4}, this phase space is sampled discretely, thereby generating a large amount of model lightcurves. We find that the simulated Stokes parameters ($\xi$), vary smoothly as a function of this parameter space, showing systematic behaviour as we vary the observed energy band, viewing angle $\chi$, pulsar inclination $\alpha$, and particle pitch angle distributions. We discuss initially, the form of the simulated Stokes parameters, how these vary as a function of parameter space and how this variation can be used to select a best fitting parameter combination (the `search algorithm'). We subsequently describe our restrictions on pulsar geometry ($(\alpha,\chi)$ combination), and discuss the inverse mapping step which locates the associated source region within the pulsar magnetosphere.\\

We note here, that  our simulations resolve pulsar phase into 200 bins, and we subsequently quote the phase extent of a single pulsar cycle as going from 0 $\rightarrow$ 1. As choice of a reference phase is arbitrary, we define phase 0 to be the phase at which the magnetic pole associated with the quoted ${\alpha}$ value would be seen by an external observer. From symmetry, phase 0.5 is the phase at which the magnetic pole at ${\alpha=180\deg-\alpha}$ would be seen. We refer to the magnetic pole at inclination $\alpha$ as pole 1 and the magnetic pole at $180\deg-\alpha$ as pole 2.

\subsection{Form and variation of simulated Stokes parameters}

The search algorithm method, in sampling such a large parameter space, inherently creates a large amount of data. It is instructive initially to discuss the overall form of the simulated Stokes parameters and to describe how they vary as a function of the phase space parameters  $\alpha, \chi,$ \pad, and \padco. Some useful systematic trends are evident which will be discussed.

\subsubsection{\pad$\:$and \padco}

It is found that similar trends exist for the variations of $\xi$ between the different \pads combinations (i.e., each of the \pads functions based on isotropic, circular, linear and Gaussian profiles all have similar $\xi(\alpha,\chi)$). We find that varying the \padcos has a larger effect on the polarisation than varying the \pad  itself. The general trend is for larger \padcos values to generate broader peaks, but the peaks are located at the same phase locations so that the overall variation of $\xi(\alpha,\chi)$ is similar for all \pad s. Given this degenerate behaviour, we will focus our analysis on the isotropic pitch angle distribution, which should reflect the overall general trends in the variation of $\xi(\alpha,\chi)$.

\subsubsection{Stokes I: Single and Double Peaks}

We have found that simulated lightcurves only have a single peak for any inclination $\alpha \lesssim 50\deg$, but a double peak profile is seen to emerge for $\alpha \gtrsim 50\deg$. The intensity of the secondary peak increases as $\alpha \rightarrow 90\deg$, where the main:secondary peak ratio varies smoothly with viewing angle. This allows for an immediate restriction in terms of predicted inclination and viewing angle for single and double peaked pulsar profiles.\\

\begin{figure*}
\vspace{-10em}
\centering
\begin{minipage}{0.49\linewidth}
\hspace{8em}
\includegraphics[viewport=110 300 500 690, width=1.0\linewidth]{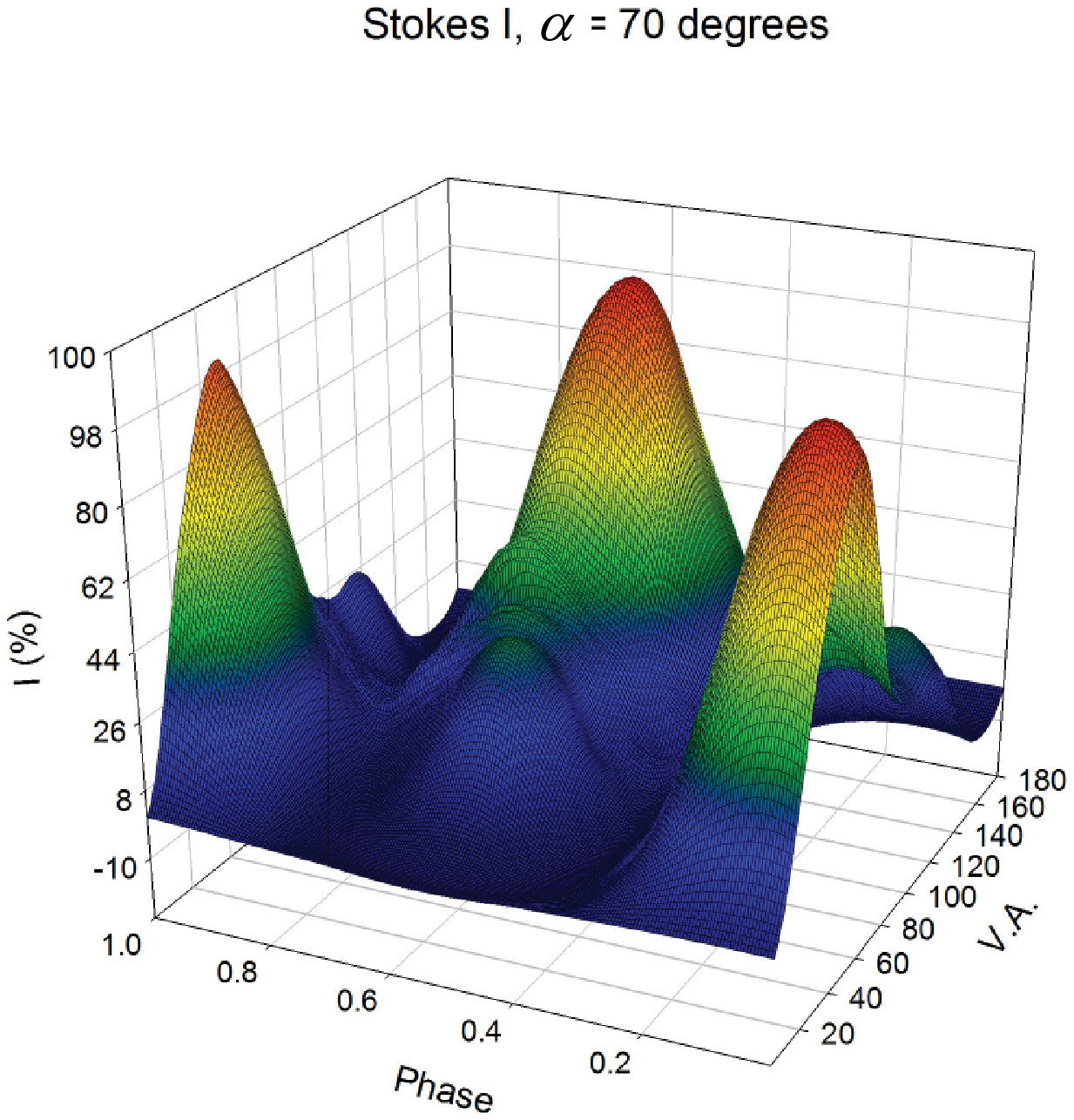}
\end{minipage}
\begin{minipage}{0.49\linewidth}
\hspace{8em}
\includegraphics[viewport=110 300 500 690, width=1.0\linewidth]{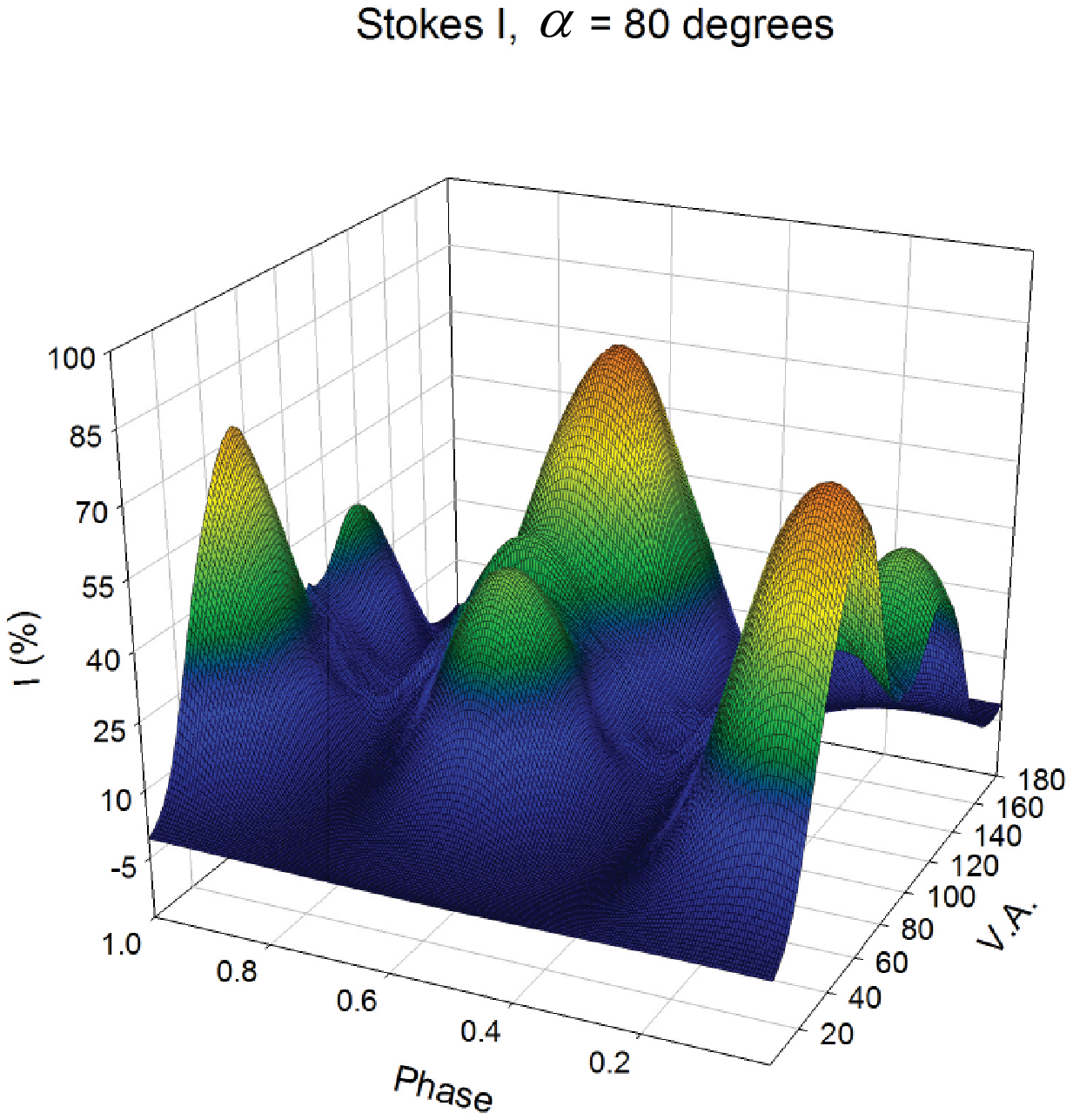}
\end{minipage}
\begin{minipage}{0.49\linewidth}
\hspace{8em}
\includegraphics[viewport=110 300 500 690, width=1.0\linewidth]{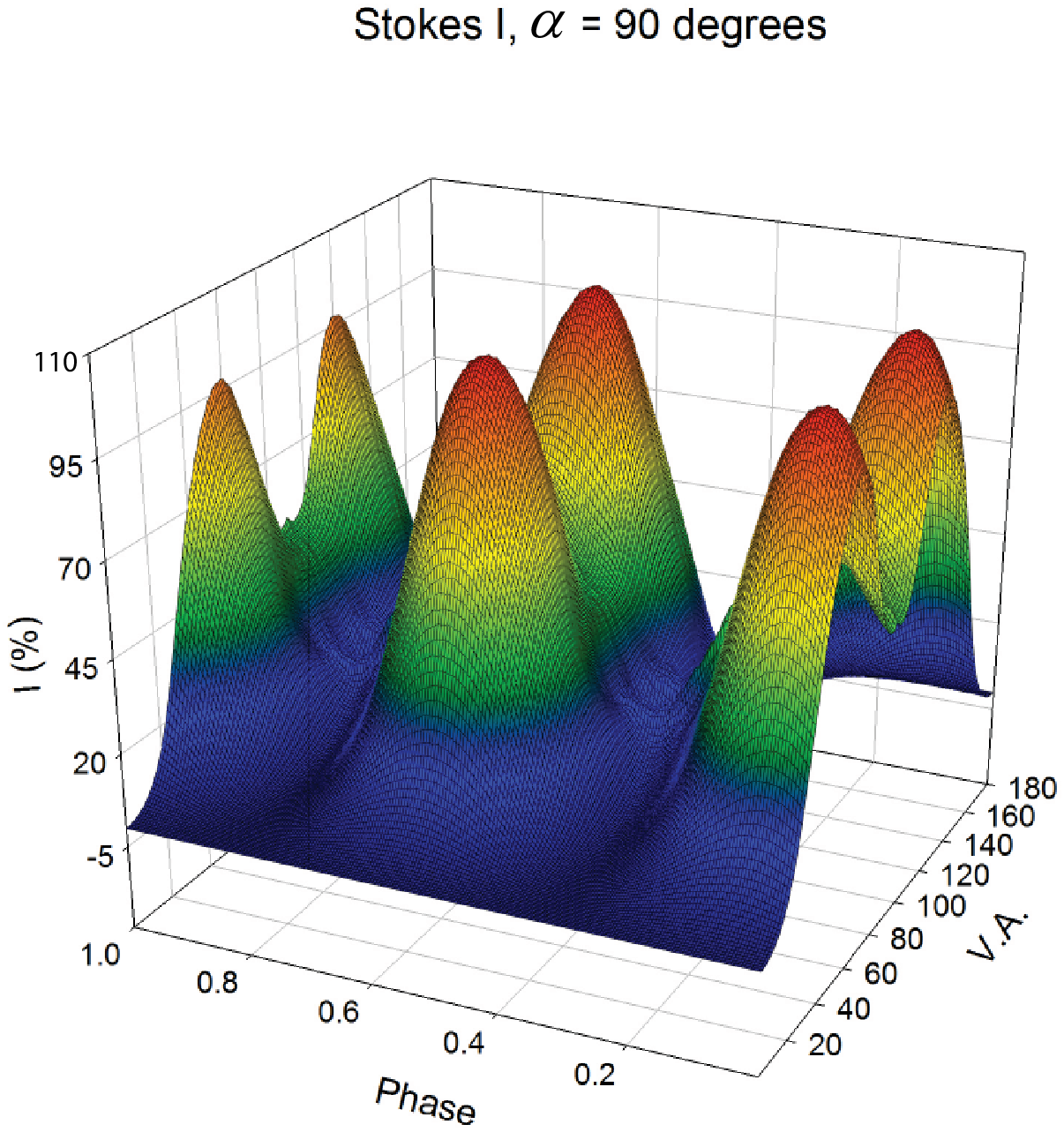}
\end{minipage}
\vspace{14em}
\caption[Primary results 4]{Relative intensity (Stokes I) at all viewing angles (V. A.) for successive pulsar inclinations. Parameter sets are (clockwise from top left) $(\alpha)= 70\deg, 80\deg, 90\deg$ with (\pad, \padco)$=$(isotropic, $20\deg$) in each case. The secondary peak increases its intensity and viewing angle extent, as $\alpha \rightarrow 90\deg$. At $\alpha=90\deg$, the `secondary peak' is of equal intensity to the main peak. In terms of restricting parameter space, this evolution shows that the observed main:secondary peak ratio of $\bm{ \sim$ $\bm 0.3}$ (for the Crab pulsar), can only be reproduced for a restricted range of viewing angles (for $\bm{\Delta \chi \sim 5\deg}$ about a central $\bm \chi_o$) for $\bm{ \alpha \gtrsim 60\deg}$}.
\label{3d-peak2}
\end{figure*}

\begin{figure}
\centering
{\includegraphics[scale=0.33,angle=-90,clip]{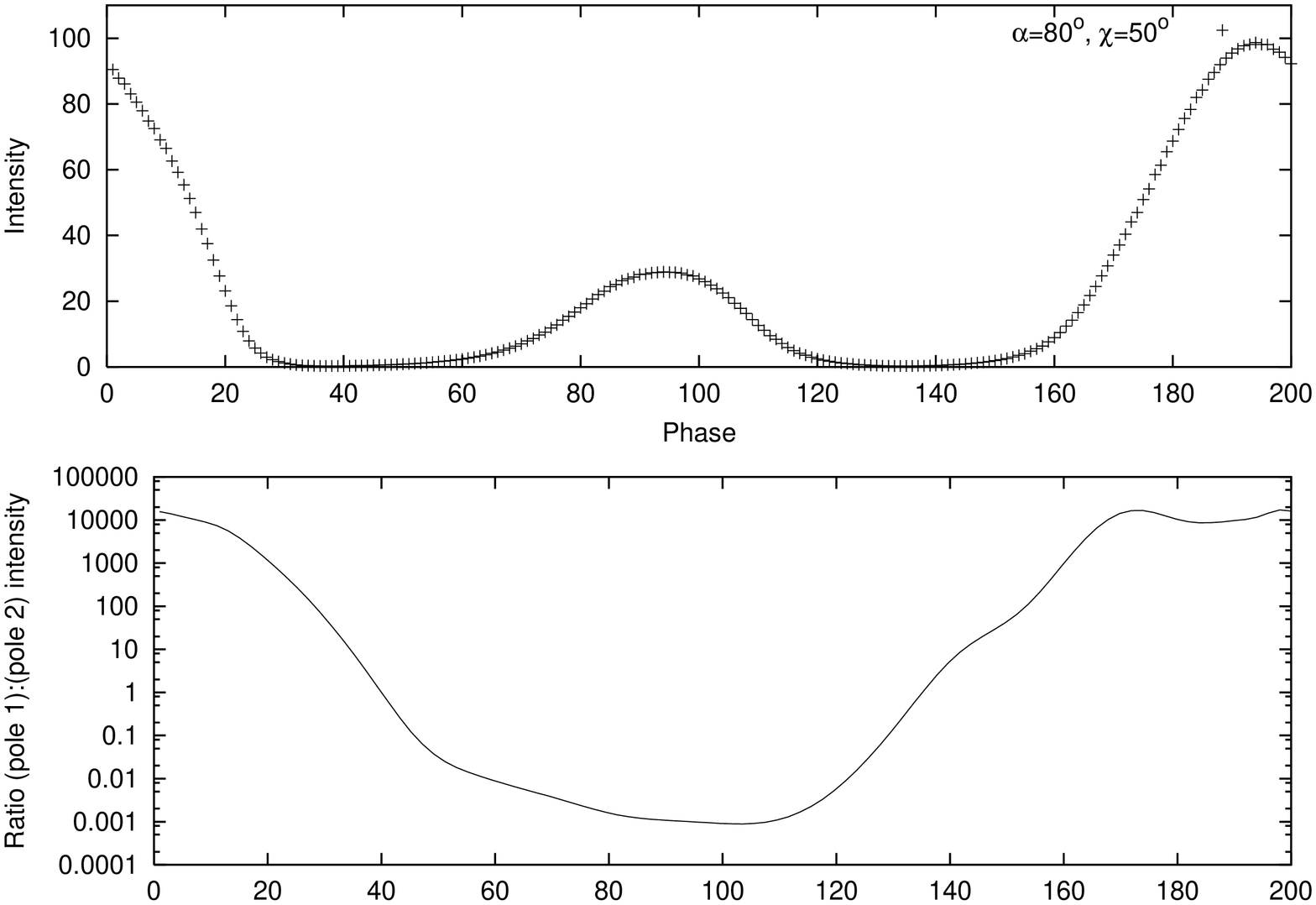}}
\hfill
{\includegraphics[scale=0.33,angle=-90,clip]{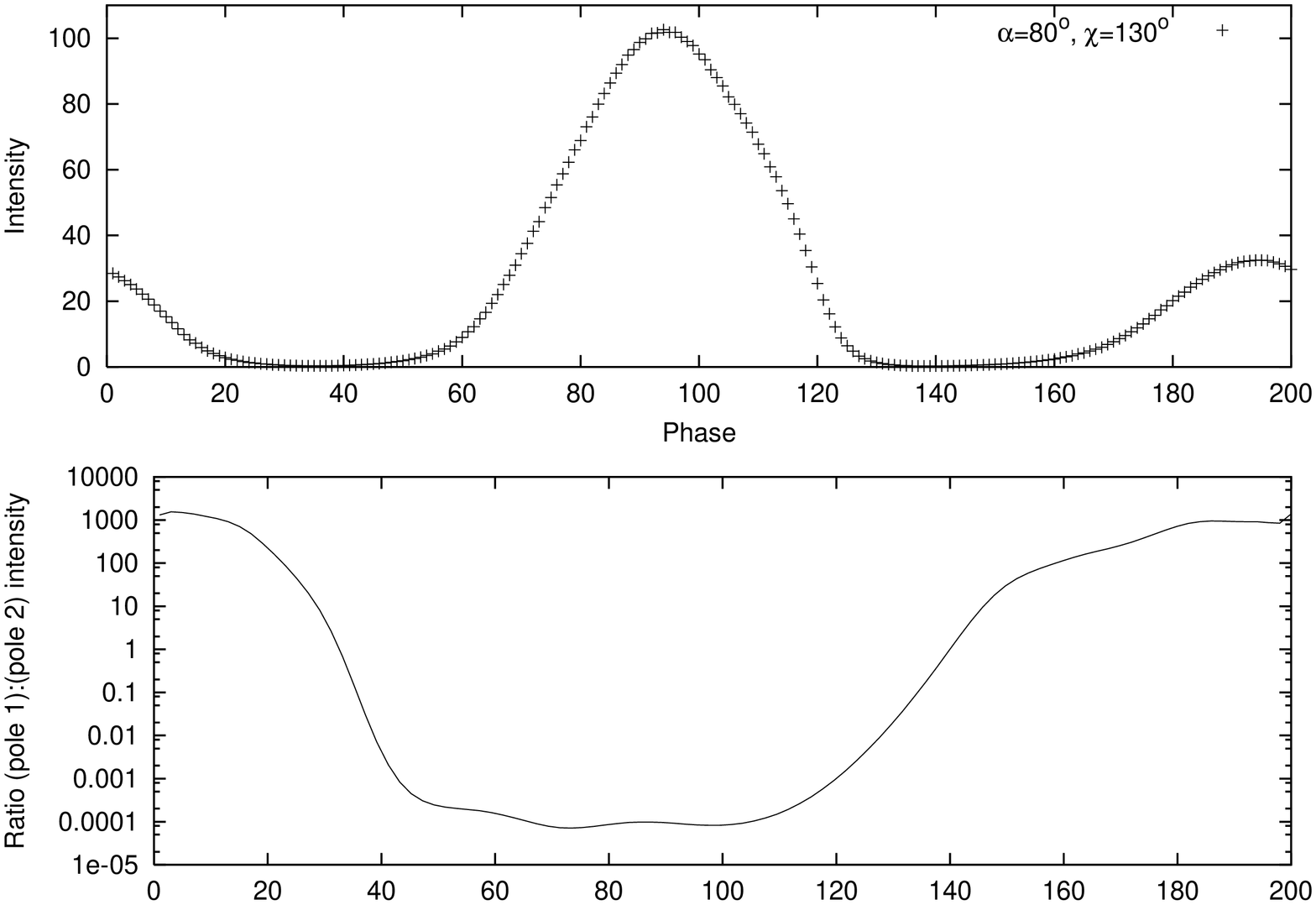}}
\caption[Primary results 6]{Intensity observed for $(\alpha)=80\deg$ showing relative relative flux contribution of emission from pole 1 and pole 2 for different viewing angles.  Here pole 1 refers to the magnetospheric regions above pole 1 and not just to the polar cap regions. Results illustrate that, depending on whether $\chi\gtlt90\deg$, different poles will contribute to main and interpulse emission.}
\label{10}
\end{figure}

Using $\xi$=I$(\bm{R},\alpha,\chi,\Phi)$, it is found that the main emission contribution to the secondary peak, comes from pole 2 if $ \chi\lesssim90\deg$, and from pole 1 if $ \chi\gtrsim 90\deg$, indicating that as $ \alpha\gtrsim 50\deg$, the secondary peak is formed from the magnetic pole {\it opposite} the pole contributing to the main peak. Figure~\ref{10} shows some phase resolved plots of the I(pole 1):I(pole 2) ratios for an inclination of $\alpha=80\deg$, at different viewing angles, which illustrates the relative contributions of emission over each pole to the final integrated lightcurve.\\

\subsection{General Observational Restrictions}

Certain features of Crab pulsar emission are clear and distinct and any successful modelling of this object should be able to recreate at least some of these main features:

\newcounter{listno5}
\begin{list}{ [\alph{listno5}] }
  {\usecounter{listno5}
  {\setlength{\rightmargin}{0.1\columnwidth}}
  {\setlength{\leftmargin}{0.1\columnwidth}}}

\item {\bf Double peaks:} The Crab pulsar exhibits a double peaked structure at all wavelengths. Only simulated dipole inclinations with $\alpha \gtrsim 50\deg$ are capable of producing double peak emission. This is our first restriction of phase space.\\

\item {\bf Relative peak intensity and bridge emission:} The double peaked structure is accompanied by the requirements of explaining the relative intensity of the peaks. A secondary peak whose intensity and extent of visibility (range of $\chi$ values over which it is visible) is a (monotonic) function of inclination $\alpha$, appears within the simulations for $\alpha \gtrsim 50\deg$. The ratio of main:secondary peak intensity of 1:0.3, is possible for a narrow range of $\chi$ associated with any $\alpha \gtrsim 60\deg$. This range of $\chi$ extends no more than $\sim 2.5\deg$ about a central $\chi=\chi_o$. We find that $\chi_o$ is a smooth function of pulsar inclination ($\chi_o=\chi_o(\alpha)$), $\chi_o$ decreasing steadily as $\alpha$ increases.\footnote{$\chi_o \gtrsim 80\deg$ for $\alpha = 60\deg$, $\chi_o \sim 47\deg$ for $\alpha=80\deg$, as can be seen in Figure~\ref{3d-peak2}.} Our second restriction can be summarised as follows: only for $\alpha \gtrsim 60\deg$ and for  viewing angles $\Delta \chi \sim 2.5\deg$ about $\chi_o=\chi_o(\alpha)$, are simulated lightcurves able to reproduce the main:secondary peak ratio of $\sim$ 1:0.3. Bridge emission is also a function of $\alpha$ where the large bridge emission seen at lower inclinations ($\alpha$), tends to disappear for the more orthogonal rotators. \\

\item {\bf Peak phase separation:} The Crab pulsar's peak separation is $\sim 0.4$ in phase. Results indicate that the simulated model is unable to produce a double peak structure separated by less than $0.4$ in phase, where in fact most viewing angles have peak separations of 0.5. In hindsight, this is a result of the inherent symmetry of the physical model, as it simulates emission from the open volume of both poles of a symmetric dipolar magnetic field structure. The influence of relativistic beaming can lead to asymmetries in pulse structures, so one would sensibly infer that a global emission model viewed at $\alpha\ne 90\deg, \chi \ne 90\deg$ should subsequently exhibit non-symmetric behaviour. It seems therefore, that the explanation for the symmetry in these model results, most likely lies in the averaging of emission over such a large volume. The large width of the peaks is also most likely due to the large volume of emission. It is hoped that the inclusion of more general physical constraints into the model, should lead to a removal of the symmetry.\\

\end{list}

\subsection{Inverse Mapping - A two pole emitter?}

\begin{figure*}
\centering
{\includegraphics[scale=0.65,angle=0,clip]{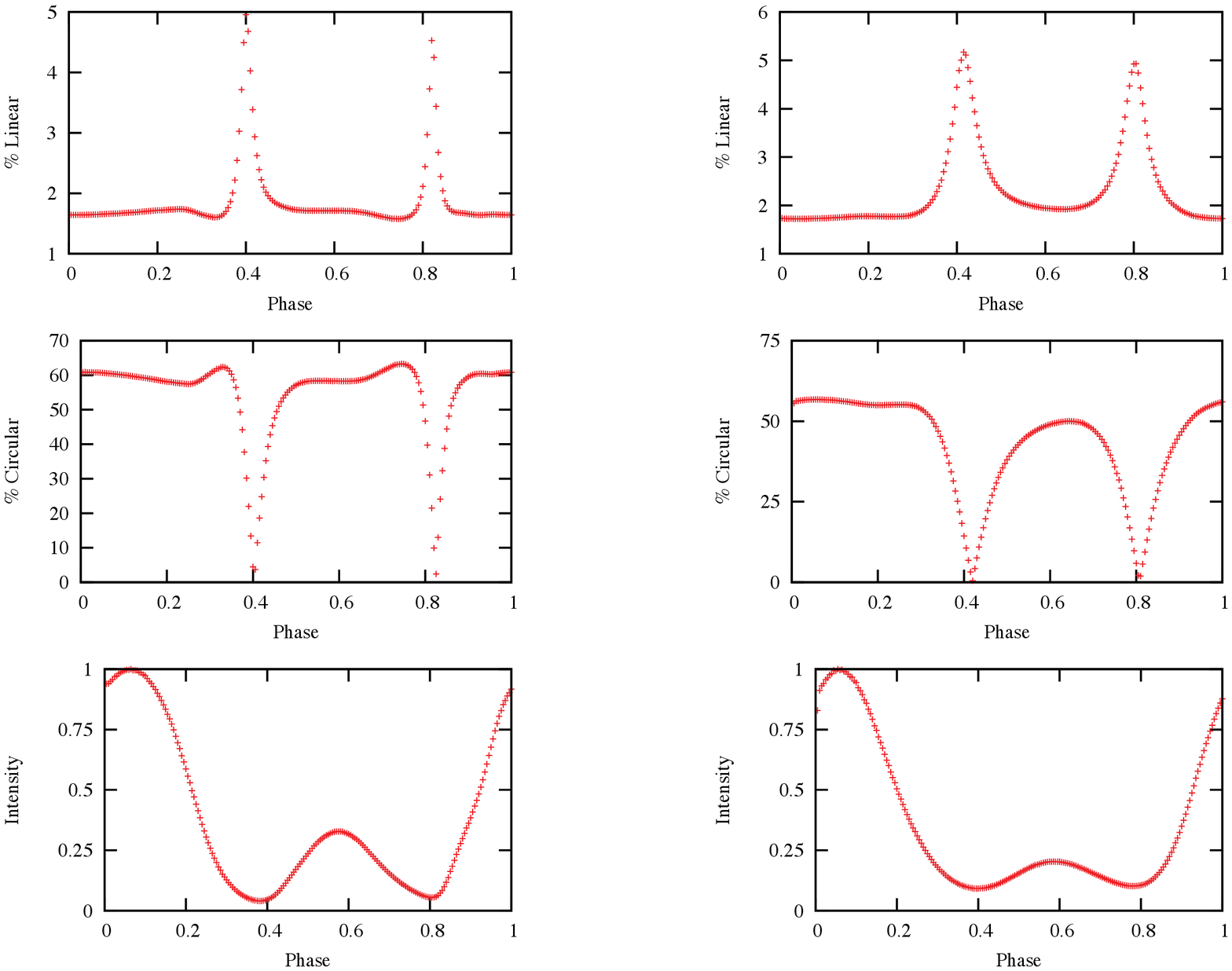}}

\hfill

\caption[Phase Plots]{Phase plots for 70$\deg$ inclination and viewing angles 45$\deg$ and 20$\deg$. We show the total optical intensity, and degree of linear/circular polarisation.  }
\label{Phase_Plots}
\end{figure*}

\begin{figure*}
\centering

{\includegraphics[scale=0.53,angle=0]{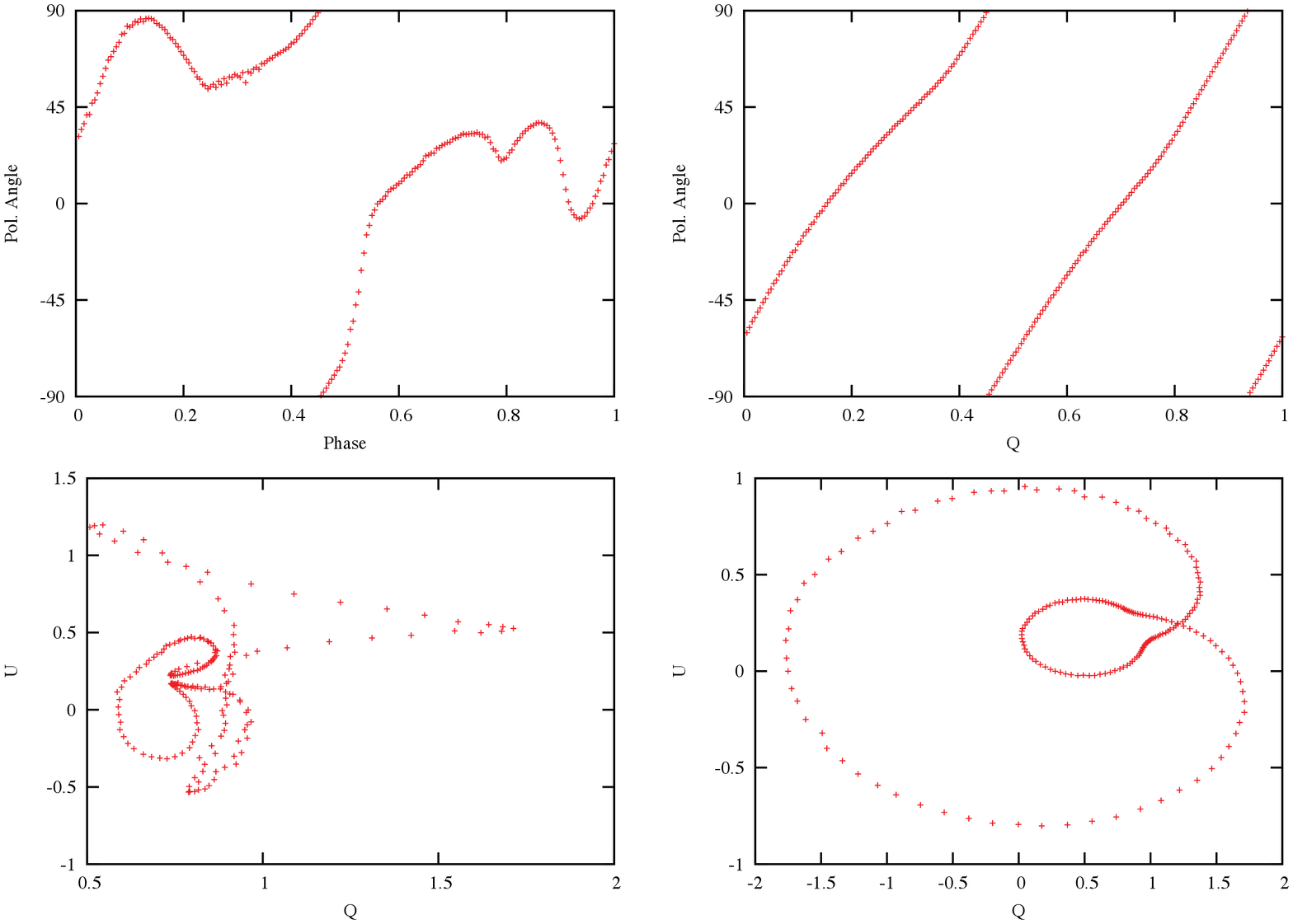}}

\caption[PA Plots]{Illustrative polarisation angle vs phase and QU plot  for 70$\deg$ inclination and viewing angles 45$\deg$ and 20$\deg$. }
\label{PA_Plots}
\end{figure*}

Our results given that the model results are capable of more successfully reproducing  lightcurve flux profiles than the associated polarisation 
profiles, inverse mapping results are given for a selected ($\alpha,\chi$) combination, which compares most favourably to the observed 
double peak structure of the Crab pulsar and the observed QU relationship. A value of $(\alpha,\chi)=(70\deg,45\deg)$ is chosen 
for (\pad, \padco)=(isotropic, $20\deg$), these values were based on a best fit in the $\chi^2$ sense between the simulated data and the \cite{SJDP1988} 
observations. In Figure \ref{Phase_Plots} we show the intensity and degree of linear and circular polarisation as a function of phase. We note that the intensity distribution is wider than normal observed - see for example \cite{2009MNRAS.397..103S} this can be explained by our initial approach of filling the open field region with radiating particles. \cite{2009MNRAS.397..103S} also showed an increase in the degree of linear polarisation happening prior to the main pulse which we see although not at exactly the same phase relationship. From Figure \ref{PA_Plots}  we see the expected swing in polarisation angle around the two peaks and QU plots morphologically similar to \cite{2009MNRAS.397..103S} and \cite{S1986}. Our preferred values of $\alpha$ and $\chi$  are broadly consistent with \cite{2004ApJ...601..479N}, which provided a robust estimate of the pulsar inclination angle.

The inverse mapping element of the search algorithm allows us to decompose light-curves, with the option of selecting different phase resolved regions (on our current resolution phase is divided into 200 bins), which we can subsequently retrace back into the \mag, isolating the originating locations of the component photons. This will localise the magnetospheric regions which contribute to emission at a particular phase, one of the main motivations for the search algorithm.\\

\begin{figure*}
\begin{minipage}[a]{1.0\linewidth}
\centering
\includegraphics[scale=0.6,angle=0]{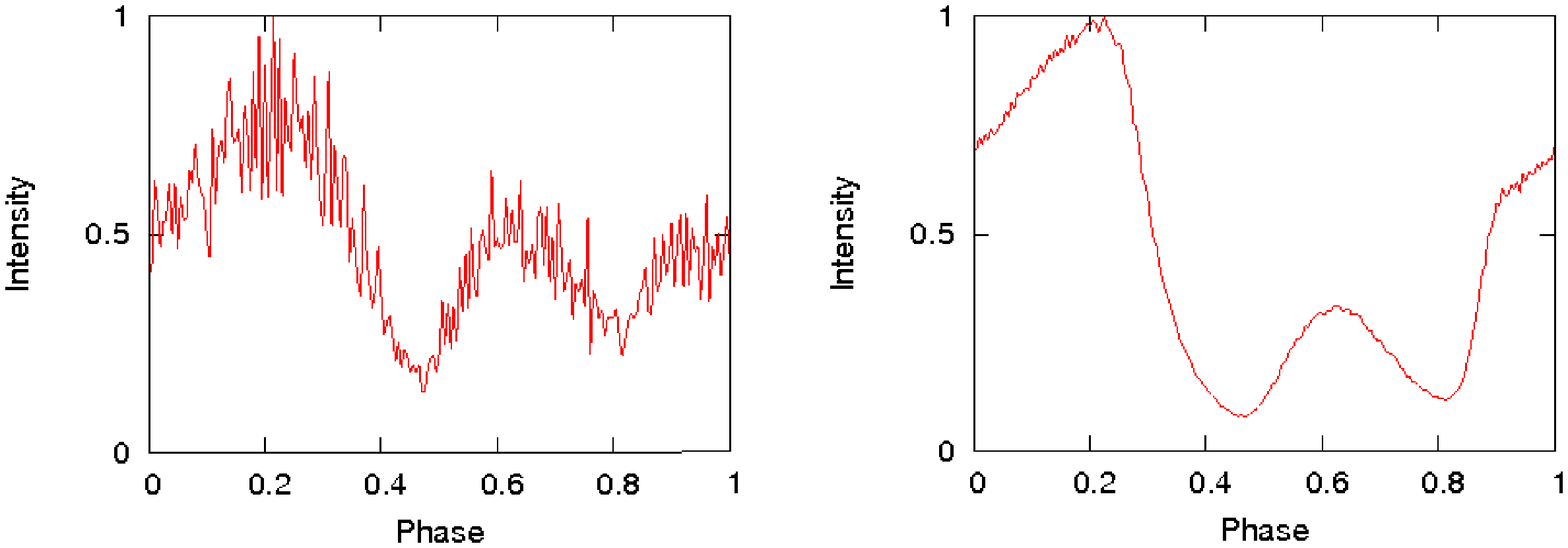}
\caption[POD1 Plots]{Phase and intensity relationship for 70$\deg$ inclination and viewing angle 45$\deg$ for two $\pad$ of 1 and 5 showing the ragged nature of the intensity for the lower pitch angles. }
\label{fig:figure1}
\end{minipage}
\hspace{0.5cm}
\label{POD1_Plot}
\end{figure*}

We try and present the results of the inverse mapping in an interpretable sense here in 2 and 3 dimensions, although the task is not an easy one. The selected $(\alpha,\chi,$\pad,\padco) combination chosen to illustrate the inverse mapping is $(70\deg, 52\deg,\:$isotropic, $20\deg$), which has a main peak at phase  $\sim 0.1$ and an interpulse at phase $\sim 0.6$. We choose specific phase regions ($\Delta\Phi=5$) centred on the arrival phase of the main peak and the inter-pulse, so as to obtain a view of the magnetospheric regions which contribute to `peak' intensity.We note however on the basis of Figure \ref{10} that the fit is a smoothly varying function of $\alpha,\chi$,\pad and \padco. In Figure \ref{POD1_Plot} we show the effect of changing the \padco showing intensity vs phase for \padco values of 1$\deg$ and 5$\deg$. \\ 

\begin{figure*}
\includegraphics[scale=0.31,clip]{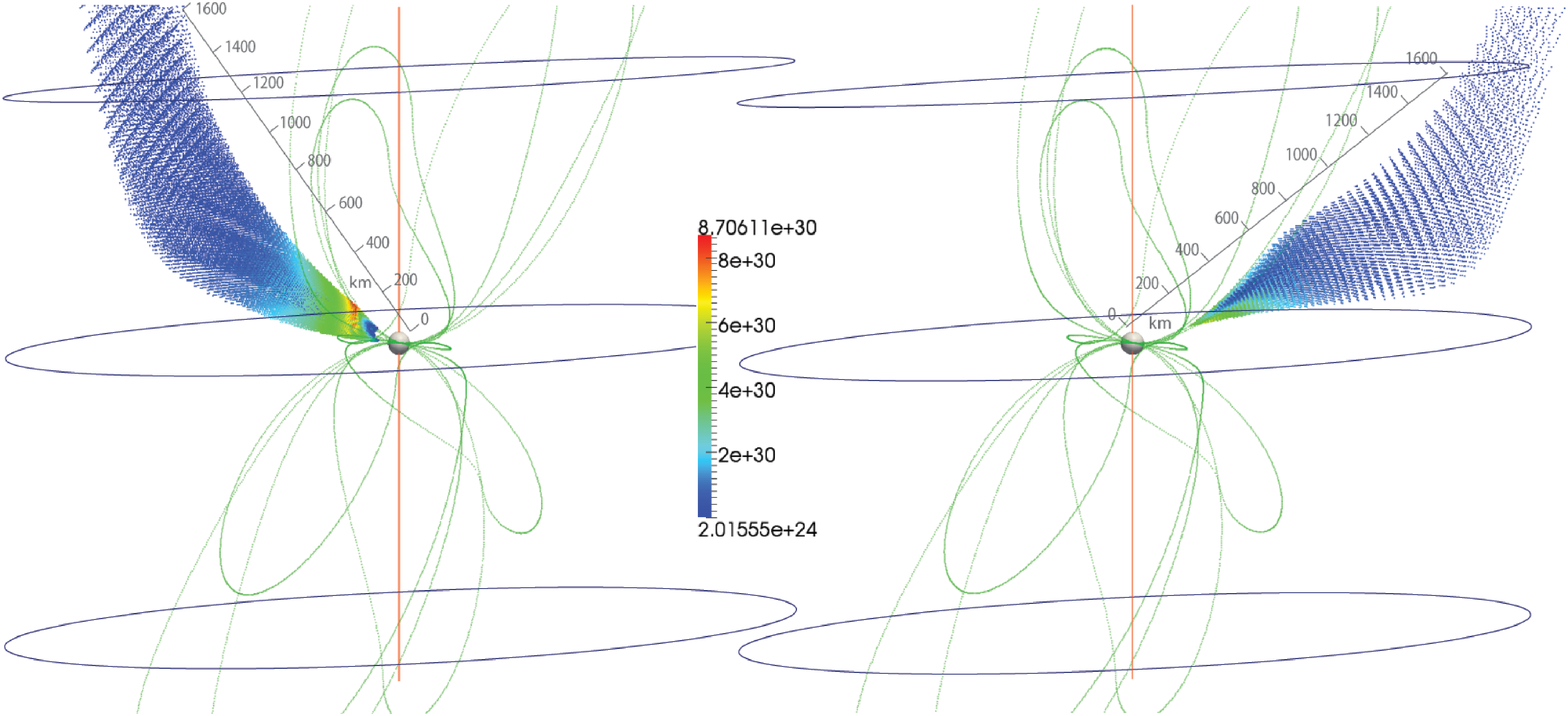}
\caption[3D representation of inverse mapped regions for main peak emission]{Three dimensional representation of the regions contributing to {\bf main peak} (left) and {\bf secondary peak} (right) emission for a model (Crab) pulsar with parameters ($\alpha,\chi$,\pad,\padco)=($80\deg,25\deg,\:$isotropic, $20\deg$). The intensity scale shows the emission contributions at the regions mapped. For reference, the circles define the light cylinder boundary and the light brown lines are the open volume/closed volume boundary. Emission is seen to be concentrated in the lower magnetosphere from a single pole.}
\label{main-peak-3d}
\end{figure*}

Carrying out the inverse mapping step for $(\alpha,\chi)=(70\deg,45\deg)$, Figure~\ref{main-peak-3d} presents a 3D colour coded intensity map of the originating location of emitted radiation seen by an observer at $\chi=52\deg$ relative to the model pulsar. Figure~\ref{main-peak-3d} (left) shows the origin of emission which composes the main peak and Figure~\ref{main-peak-3d} (right) shows the origin of emission composing the interpulse (as an aid to perspective, the circles define the light cylinder distance and the brownish field lines are the boundary of the open volume/closed volume region within the pulsar magnetosphere). Immediate points of note, are that the main peak is primarily composed of emission from one pole (which we term pole 1), while the interpulse is composed of emission from the opposite pole (here defined as pole 2). Also, the most intense emission is located in tightly constrained volumes in the inner magnetosphere.\\

It is immediately apparent that these results show localised emission regions and clearly indicate a two pole emission model. Figure~\ref{main-peak-3d} illustrates rather dramatically, that the first order assumption of a global truncated power law synchrotron emission model (even for a relatively wide pitch angle distribution of $20\deg$) can produce a rather limited spatial extent of maximum emission within the pulsar magnetosphere, while matching certain features observed in the Crab pulsar profile. The main location of emission is well away from the standard outer gap location and closer to the polar cap centre, but a closer analysis (described subsequently) yields that the emission is at distances of $\sim  0.2R_{LC}$ from the polar cap, further from the polar cap surface than polar cap models generally estimate. \\

In an independent analysis, we attempt to localise relative emission contribution as a function of magnetospheric location. We subdivide the magnetospheric volume into concentric spherical shells of width $\Delta\rho$ ($\rho$ being the spherical radius), and also into coaxial cylindrical shells of thickness $\Delta r$ ($r$ in cylindrical coordinates). We integrate emission from all points within these shells to obtain a radial profile (spherically and cylindrically respectively) of absolute contributions to the total emission. The spherical profile is presented in the left hand panels of Figures~\ref{13} and \ref{14} with the cylindrical contributions shown on the right. Figure~\ref{13} shows emission contributions to the main emission peak at phase $\sim 0.1$, Figure.~\ref{14} shows emission contributions to the secondary peak at phase $\sim 0.6$. The top panels in these figures show the total emission from over both magnetospheric poles, the middle panels show emission regions associated with pole 1\footnote{Pole 1 corresponds to the magnetospheric region above the 
magnetic pole at an inclination $\alpha$} only and the bottom panels show emission from pole 2\footnote{Pole 2 refers to emission
associated with the pole at inclination $180\deg + \alpha$.} only.\\

From the previous figures, certain trends are apparent. Pole 1 emission dominates for the main pulse and pole 2 emission for the secondary, each pole having distinct emissivity trends, both radially (cylindrically) and spherically. The poles dominating emission tend to have their maximum contributions located at very small spherical and cylindrical ($\sim 0.2 R_{LC}$) radii, whereas the poles which contribute less to the total emission are located further from the star. The high peak in the emissivity, may be due to the influence of higher local charge densities lower down in the magnetosphere. Generally, the collective interpretation is that the main contribution to emission is very localised closer to the neutron star surface\footnote{The poles contributing less to the emission of a given peak also seem to have their emission localised, but at a greater distance from the star.} than to the light cylinder. This tends to agree with models attempting to explain optical emission through a localised effect such as the slot gap.  Indeed our results are consistent with the emission altitude predicted by \cite{2010MNRAS.405..509D}. Furthermore we note that the slot-gap models primarily localise emission in the transverse (co-latitude) direction whereas we also restrict emission in altitude.

\subsection{Conclusion}

Optical polarisation studies can be used to determine the local geometry of the emission region. In particular from this work we have
\begin{itemize}
\item A simple synchrotron model for the emission gives reasonable agreement with observations.
\item A prediction that the emission is low in the magnetosphere at an altitude in the region 30-40 stellar radii, that is away from both the polar cap and likely outer gap regions. 
\item The linear polarisation peaks on the rising edge of the main pulse, consistent with the findings of \cite{SJDP1988,2009MNRAS.397..103S}.
\item The radiation for the most part is circularly polarised with an interplay between  linear and circular polarisation on the rising edge of the main pulse.  On this basis observations of pulsed circular polarisation from optical pulsars would provide a significant geometrical restrict on pulsar parameters although we accept more detail work is needed of more realistic emission zones. 
\item Pulse widths which are significantly greater than those observed although this probably stems from our requirement to fill to open field region with an emitting plasma.
\item Due the inherent symmetry of this model we do not see any significant bridge emission at the preferred orientation and viewing angle.

\end{itemize}

Our future work will entail restricting emissions to localities around the last open field lines thereby removing the inherent symmetry existing within the model. In this way we hope to able to make firmer predictions in relation to the optical emission specifically and more generally comment on correlations between the optical and radio emission \cite{2009MNRAS.397..103S}.  As well as the emission location and pulsar orientation our approach can also restrict the \pad and \padco. We also intend to make more detailed comparison between these predictions and measurements of the pulsed circular polarisation. To-date there have been no measurements of optical circular polarisation from any pulsar.  A new instrument, the Galway Astronomical Stokes Polarimeter (GASP), \cite{2010EPJWC...505003K} has this capability and observations are planned to observe the Crab pulsar in late 2011.


\section*{Acknowledgments}
The authors would like to thank the Higher Education Authority Programme for Research in Third Level Institutions (PRTLI) Cosmogrid project and National University of Ireland's Millenium Fund for financial support for POC for the duration of this work and the PRTLI 4 E-Inis project for financial support for JMcD for the duration of this work. Finally we would also like to thanks the anonymous referee whose comments made significant improvements to an earlier version of this paper.

\pagebreak

\appendix

\section[]{Synchrotron polarimetry for a truncated power law spectrum of particles}
\label{appendix_1}

The effect of the $truncated$ power law energy spectrum results in the functions modifying the observed Stokes parameters (i.e.,~$\mathcal{J}_n(x), \mathcal{L}_n(x)$ and $\mathcal{R}_n(x)$), needing to be evaluated between upper ($x_1$) and lower ($x_2$) limits of the parameter `x', where $x=f/f_c$. This introduces an explicit dependency on the underlying particle energy limits, $E_1$ and $E_2$. It is the functional form of these x-factor limits and their variation, which dictates the resultant polarised emissivity spectrum for synchrotron radiation. 

As described by GLW74, the polarised emissivity can be investigated by defining and introducing unique correction factors, $C^{(i)}(x_l,x_2)$ for each of the Stokes parameters, where:

\begin{equation}
\parbox{0.88\columnwidth}{
\begin{eqnarray*}
 C^{(1)}(x_1,x_2) & = & \left[ \mathcal{J}_{(\gamma+1)/2} \right]_{x_2}^{x_1} / \mathcal{J}_{(\gamma+1)/2}\\
 C^{(2)}(x_1,x_2) & = & \left[ \mathcal{L}_{(\gamma+1)/2} \right]_{x_2}^{x_1} / \mathcal{L}_{(\gamma+1)/2}\\
 C^{(3)}(x_1,x_2) & = & \frac{\left[ \mathcal{R}_{(\gamma/2+1)}+(1+g(\theta))(\mathcal{L}_{\gamma/2}-\frac{1}{2}
                               \mathcal{J}_{\gamma/2}) \right]_{x_2}^{x_1}}{
                        \left[ \mathcal{R}_{(\gamma/2+1)}+(1+g(\theta))(\mathcal{L}_{\gamma/2}-\frac{1}{2}
                               \mathcal{J}_{\gamma/2}) \right]}
\end{eqnarray*}} 
\end{equation}
with:
\begin{equation}
 C^{(i)}(x_l,x_2)  = C^{(i)}(x_l,0) - C^{(i)}(x_2,0), \quad i=1,2,4
\end{equation}
where,
\begin{equation}
\label{xl}
x_l = \rm{min}(x_1,x^\prime_1)
\end{equation}
where $x^\prime_1$ is the frequency limit cutoff x-factor as discussed in section 3.2.\\

The correction factors encode the full details of the effects due to the truncated nature of the particle spectrum. They include the explicit dependency on $x_1$ and $x_2$ and further, they encode the physical effect of the frequency limit cutoff energy by replacing the limit $x_1$ with, $x_1'$ whenever $E_1<E_1'$. 

Once $x_a<x^{(i)}_b$, the four fiducial x-factors defining the end-points and transition-points are $x_l<x_a<x^{(i)}_b<x_u$. Using asymptotic approximations to the PLEs (Power Law Expansions) of GLW74, the variation of the correction factors can be expressed as follows, where $i=1,2$ and $j=4$:

\noindent
\parbox{0.25\columnwidth}{
$
  \setlength{\arraycolsep}{0.0em}
  \begin{array}{cc}   
               \\ \\ \\ C^{(i)}(x_l,x_2) & {}= \\ \\ \\\\ \\                
  \end{array}
  \left\{
  \begin{array}{c}   
               \\ \\ \\  \\ \\ \\\\ \\                
  \end{array}
  \right.
  \setlength{\arraycolsep}{5pt}
$}
\parbox{0.8\columnwidth}{
\setlength{\arraycolsep}{0.0em}
\begin{eqnarray}
\label{full_cfs1}
  & & \left[1-\left(\frac{E_1x_i^*}{E_2x_1}\right)^{(3\gamma-1)/3}\right]
     \left[\frac{x_1^3}{x_1^{*2}x^{(i)}_b(\gamma)}\right]^{(3\gamma-1)/6}  R1 \nonumber \\
  & & \left[1-\left(\frac{E_1}{E_2}\right)^{(3\gamma-1)/3}\right]
                      \left[\frac{x_1}{x^{(i)}_b(\gamma)}\right]^{(3\gamma-1)/6}
                        \hs  R2 \nonumber \\
  & & 1-\left(\frac{E_1}{E_2}\right)^{(3\gamma-1)/3}
     \left[\frac{x_1}{x^{(i)}_b(\gamma)}\right]^{(3\gamma-1)/6}
     \hs R3 
\end{eqnarray}
}

\noindent
\parbox{0.25\columnwidth}{
$
  \setlength{\arraycolsep}{0.0em}
  \begin{array}{cc}   
               \\ \\ \\ C^{(j)}(x_l,x_2) & {}=  \\ \\ \\\\ \\                
  \end{array}
  \left\{
  \begin{array}{c}   
               \\ \\ \\  \\ \\ \\\\ \\                
  \end{array}
  \right.
$}
\parbox{0.8\columnwidth}{
\setlength{\arraycolsep}{0.0em}
\begin{eqnarray}
\label{full_cfs2}
 &  & \left[1-\left(\frac{E_1x_i^*}{E_2x_1}\right)^{\gamma}\right]
     \left[\frac{x_1^3}{x_1^{*2}x^{(4)}_b(\gamma)}\right]^{\gamma /2}
     \hs R1  \nonumber \\
 &  & \left[1-\left(\frac{E_1}{E_2}\right)^{\gamma}\right]
     \left[\frac{x_1}{x^{(i)}_b(\gamma)}\right]^{(3\gamma-1)/6}
     \hs   R2 \nonumber \\
 &  & 1-\left(\frac{E_1}{E_2}\right)^{\gamma}
     \left[\frac{x_1}{x^{(4i)}_b(\gamma)}\right]^{\gamma/2} 
     \hs \hs \hs \hs R3
\end{eqnarray}}
\noindent
where:
\begin{equation}
x_l=\rm{Min}(x_1,{x_1^{\prime}});\hspace{2em} {x_1^{\prime}}=\frac{x_1^3}{x_1^{*2}};
 \hspace{2em} x_1^*=\frac{2}{3}\left( \frac{E_o}{E_1 \sin\theta} \right)  \nonumber
\end{equation}
\noindent with $x_1 \gtlt x_1^\prime$ as $x_1 \gtlt x_1^*$. The correction factors have three major regions (referred to as $R1, R2$ and $R3$ above) with asymptotically distinct behaviour. The boundaries are defined by the transition x-factors, which also define the frequency transition points in the photon spectrum. 
\begin{eqnarray*}
R1: & & \frac{E_1}{E_2}x_1^*<x_1<x_1^*\\
R2: & & x_1^*<x_1<x_b^{(4)}(\gamma)    \\
R3: & & x_b^{(4)}(\gamma)<x_1<{\left(\frac{E_2}{E_1}\right)}^2x_b^{(4)}(\gamma) \\
R1: & & \frac{f}{f_B \sin^2 \theta} \frac{E_o}{E_1} < f < \frac{f}{f_B \sin^2 \theta} \frac{E_o}{E_1} \\
R2: & & \frac{f}{f_B \sin^2 \theta} \frac{E_o}{E_1} < f < \frac{3}{2} f_b \sin \theta \left( \frac{E_1}{E_o} \right)^2  x_b^{(i)}(\gamma)  \\
R3: & & \frac{3}{2} f_b \sin \theta \left( \frac{E_1}{E_o} \right)^2  x_b^{(i)}(\gamma) < f < \left( \frac{E_2}{E_1} \right)^2 f^{(i)}_b(\gamma) 
\end{eqnarray*}

The photon spectrum corresponding to the transition x-factors are as follows:

\noindent
\parbox{0.2\columnwidth}{
\setlength{\arraycolsep}{0.0em}
$
  \begin{array}{cc}   
               \\ \\ \\ S^{(i)}(f) & {}=  \\ \\ \\\\ \\                
  \end{array}
  \left\{
  \begin{array}{c}   
               \\ \\ \\  \\ \\ \\\\ \\                
  \end{array}
  \right.
$}
\hspace{-1cm}
\parbox{0.7\columnwidth}{
\begin{eqnarray}
\setlength{\arraycolsep}{0.0em}
\label{full_fcfs1}
   & & (f_B \sin \theta) \hspace{3mm} \left(\frac{f}{f_B \sin \theta} \right)^\gamma \hs  R1 \nonumber \\
   & & (f_B \sin \theta) \hspace{3mm} \left(\frac{f}{f_B \sin \theta} \right)^{1/3}\left(\frac{E_1}{E_0}\right)^{-(3\gamma-1)/3} \hs R2 \nonumber \\
   & & (f_b \sin \theta )^{(\gamma+1)} \left(\frac{f}{f_B \sin \theta}\right)^{-(\gamma-1)/2} \hs R3
\end{eqnarray}
}

\noindent
\parbox{0.2\columnwidth}{
\setlength{\arraycolsep}{0.0em}
$
  \begin{array}{cc}   
               \\ \\ \\ S^{(j)}(f) & =  \\ \\ \\\\ \\                
  \end{array}
  \left\{
  \begin{array}{c}   
               \\ \\ \\  \\ \\ \\\\ \\                
  \end{array}
  \right.
$}
\parbox{0.8\columnwidth}{
\setlength{\arraycolsep}{0.0em}
\begin{eqnarray}
\label{full_fcfs2}
  & & (f_B \sin \theta)^{(\gamma +1)/2} \left(\frac{f}{f_B \sin \theta} \right)^\gamma \hs R1 \nonumber \\
  & & (f_B \sin \theta)  \left(\frac{E_1}{E_0}\right)^{-\gamma} \hs R2 \nonumber \\
  & & (f_B \sin \theta)  \left(\frac{f}{f_B \sin \theta}\right)^{-\gamma/2} \hs R3
\end{eqnarray}}
where $i=1,2$ and $j=4$, with $S^1, S^2, S^4$ representing Stokes $I,Q,V$. 

There is one other possible spectral form which occurs if $x_1^* > x^{(i)}_b(\gamma)$ and in this case the photon spectrum is composed of a single transition point, designated $\tilde{f}_a$. One now has x-factor divisions, where $x_l<\tilde{x}_a<x_{u}$. 

For $x_{lo} < x_1 < \tilde{x}_a$, part of the particle population is excluded from contributing radiation at the frequency of interest, because their fundamental emission frequency is too high. We label this region $R1_b$, as it corresponds directly to region $R1$ of the previous case (where $x_1^* > x^{(i)}_b(\gamma)$). The range where $\tilde{x}_a<x_1<x_{hi}$, we label region $R3_b$. In this case the limits $(x_l,x2)$ are wide enough apart, so that we can relax to the classical scenario where the limits are approximated to $(0,\infty)$ respectively which corresponds to region $R3$ of the previous case. 

If $x_1^* > x^{(i)}_b(\gamma)$, the spectral form of the correction factors and resultant emissivity in regions $R1_b$ and $R3_b$, is identical to the spectral variation in region $R1$ and $R3$ shown in relations \ref{full_cfs1} - \ref{full_fcfs2}. The major effect of having $x_1^* > x^{(i)}_b(\gamma)$, is to have a single internal transition point ($\tilde{x}_a$) differing from both $x_a$ and $x^{(i)}_b(\gamma)$, which subsequently divides the spectrum into two regions as follows:
\begin{eqnarray*}
R1_b: & & \frac{E_1}{E_2}x_1^*<x_1<\tilde{x}_a\\
R3_b: & & \tilde{x}_a<x_1<{\left(\frac{E_2}{E_1}\right)}^2x_b^{(4)}(\gamma) \\
R1_b: & & \frac{f}{f_B \sin^2 \theta} \frac{E_o}{E_1} < f < \frac{f_B}{\sin \theta} f^{(4)}_b(\gamma) \\
R3_b: & & \frac{f_B}{\sin \theta} f^{(4)}_b(\gamma) < f < \left( \frac{E_2}{E_1} \right)^2 f^{(i)}_b(\gamma).
\end{eqnarray*}


\begin{figure*}
\includegraphics[scale=1.4,clip]{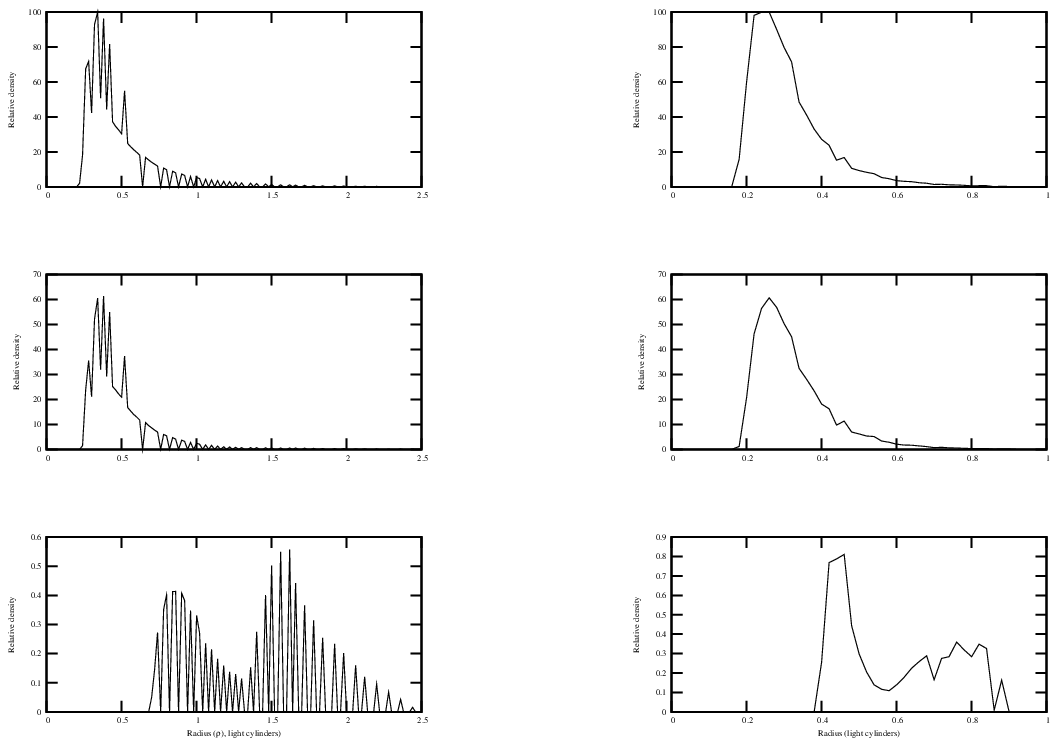}
\caption[Primary results 10]{Relative integrated emission strength within the pulsar magnetosphere for a phase extent of 0.025, centred on the {\bf main peak} (at phase $\sim 0.05$), for simulation parameters ($\alpha,\chi$, \pad, \padco) = ($80\deg, 25\deg$, isotropic, $20\deg$). Left panels: An integration of emission carried out within spherical shells concentric with the neutron star, each shell of thickness $\Delta \rho = R_{LC}/50$. Right panels: Integration of emission using co-axial cylindrical shells of thickness $\Delta r = R_{LC}/50$. Bottom panels show the location of emission associated with pole 2 only, middle panels from pole 1 only and top panels from both poles (see text for definition of `poles' and a further explanation).}
\label{13}
\end{figure*}

\label{lastpage}

\begin{figure*}
\includegraphics[scale=1.4,clip]{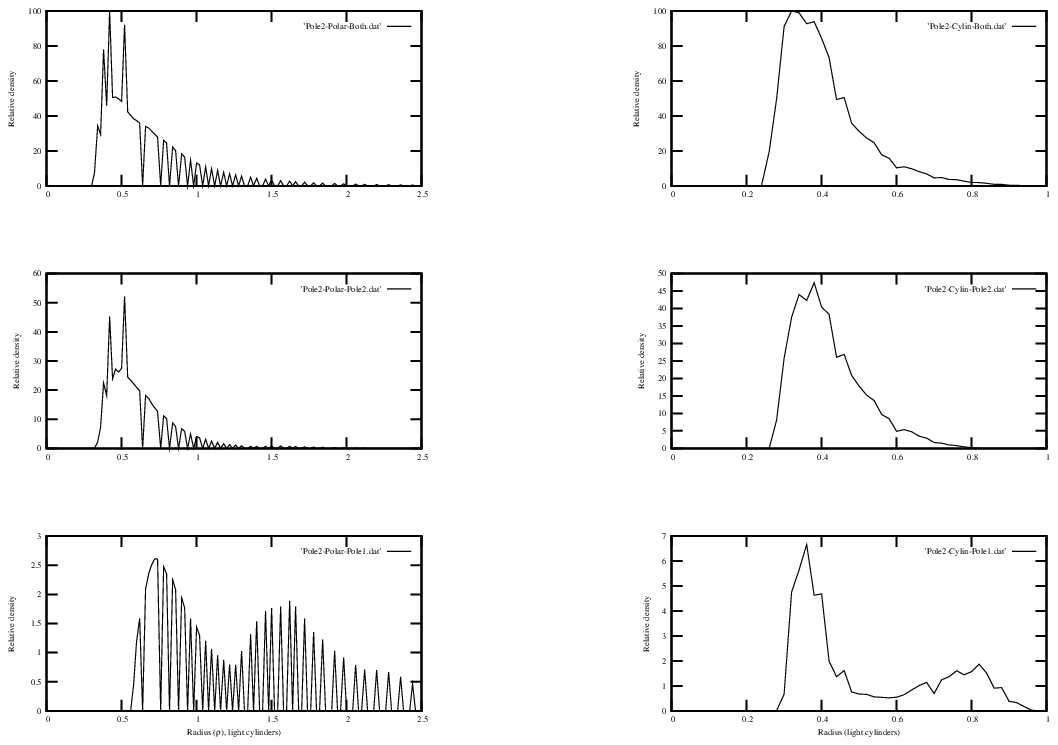}
\caption[Primary results 11]{(As figure~\ref{13} but for secondary peak). Relative integrated emission strength within the pulsar magnetosphere for a phase extent of 0.025 centred on the {\bf secondary peak} (at phase $\sim 0.55$), for simulation parameters ($\alpha,\chi$, \pad, \padco) = ($80\deg, 25\deg$, isotropic, $20\deg$). Left panels: An integration of emission carried out within spherical shells concentric with the neutron star, each shell of thickness $\Delta \rho = R_{LC}/50$. Right panels: Integration of emission using co-axial cylindrical shells of thickness $\Delta r = R_{LC}/50$.}
\label{14}
\end{figure*}

\end{document}